\documentclass[twocolumn,times]{aastex63}

\newcommand{\Msun}{\rm ~M_{\odot}}
\newcommand{\Lsun}{\rm ~L_{\odot}}

\newcommand{\NH}{\rm N_2H^+}

\newcommand{\kms}{\rm km~s^{-1}}
\newcommand{\delV}{\delta V_{HCN}}

\received{}
\revised{}
\accepted{}
\published{}
\submitjournal{}
\reportnum{}

\shorttitle{infall}
\shortauthors{M. Kim et al.}

\begin{document}
\title{Gas infalling motions in the envelopes of Very Low Luminosity Objects}

\correspondingauthor{Chang Won Lee}
\email{cwl@kasi.re.kr}

\author[0000-0002-1408-7747]{Mi-Ryang Kim}
\affiliation{Korea Astronomy and Space Science Institute, 776, Daedeokdae-ro, Yuseong-gu, Daejeon 34055, Korea}

\author[0000-0002-3179-6334]{Chang Won Lee}
\affiliation{Korea Astronomy and Space Science Institute, 776, Daedeokdae-ro, Yuseong-gu, Daejeon 34055, Korea}
\affiliation{University of Science and Technology, Korea (UST), 217 Gajeong-ro, Yuseong-gu, Daejeon 34113, Korea}

\author[0000-0002-1369-0608]{Maheswar, G}
\affiliation{Indian Institute of Astrophysics, II Block, Koramangala, Bengaluru 560 034, INDIA}

\author[0000-0002-2885-1806]{Philip C. Myers}
\affiliation{Center for Astrophysics, Harvard \& Smithsonian, 60 Garden Street, Cambridge, MA 02138, USA} 

\author[0000-0003-2011-8172]{Gwanjeong Kim}
\affiliation{Nobeyama Radio Observatory, National Astronomical Observatory of Japan, National Institutes of Natural Sciences, Nobeyama, Minamimaki, Minamisaku, Nagano 384-1305, Japan}

\begin{abstract}
We present the results of a single dish survey toward 95 VeLLOs in optically thick (HCN 1-0) and thin ($\NH$ 1-0) lines performed for the purpose of understanding the physical processes of inward motions in the envelopes of the VeLLOs and characterizing their true nature. The normalized velocity differences ($\delV$) between the peak velocities of the two lines were derived for 41 VeLLOs detected in both lines. The $\delta V$ distribution of these VeLLOs is found to be significantly skewed to the blue, indicating the dominance of infalling motions in their envelopes. The infall speeds were derived for 15 infall candidates by using the HILL5 radiative transfer model. The speeds were in the range of 0.03 $\kms$ to 0.3 $\kms$, with a median value of 0.16 $\kms$, being consistent with the gravitational free-fall speeds from pressure-free envelopes. The mass infall rates calculated from the infall speeds are mostly of the order of $10^{-6} \Msun~yr^{-1}$ with a median value of $\rm 3.4 \pm 1.5 \times 10^{-6} \Msun~yr^{-1}$. These are found to be also consistent with the values predicted with the inside-out collapse model and show a fairly good correlation with the internal luminosities of the VeLLOs. This again indicates that the infall motions observed toward the VeLLOs are likely to be due to the gravitational infall motions in their envelopes. Our study suggests that most of the VeLLOs are potentially faint protostars while two of the VeLLOs could possibly be proto-brown dwarf candidates.
\end{abstract}

\section{Introduction}
Dense molecular cores that are bright in submillimeter continuum emission but are soon to form a protostar are generally classified as starless (or prestellar) cores and cores with a detectable infrared source are classified as protostellar cores. The evolution of a starless core depends on the interplay between the inward force of gravity and the outward push of the internal pressure inside the core. Thus, the internal dynamics of the core may determine whether the core becomes static, oscillatory, expand or undergo gravitational collapse \citep{2003ApJ...586..286L,2010ApJ...721..493B,2011ApJ...734...60L,2015MNRAS.446.3731K}. 

When the inward force of gravity wins over the outward push of the internal pressure, the starless core begins gravitational contraction resulting in the movement of material towards its center. Thus, the cores that show infall motions are most likely to form a star in the future. The presence of such infalling motions can be inferred by conducting spectroscopic observations of optically thick and thin molecular lines and by looking for what is known as the ``the spectral infall asymmetry'' \citep[e.g., ][]{1997ApJ...489..719M,1998ApJ...504..900T, 1999ApJS..123..233L}. The infall asymmetry is identified with a combination of a double-peaked feature in an optically thick line where the blue peak is brighter than the red peak (such as HCN 1-0 and CS 2-1) and a single-peaked feature in an optically thin line (such as $\NH$ 1-0) \citep[e.g., ][]{1999ApJS..123..233L}. Survey observations of infall motions carried out for starless (\citealt{1999ApJS..123..233L,2006A&A...455..577T,2007ApJ...664..928S,2013ApJ...777..121S}) and protostellar cores (\citealt{1997ApJ...489..719M,1999ApJ...520..223P,2005ApJ...628L..57W,2016ApJ...833...97K}) have shown that the inward motions are a fundamental physical process that occurs in the formation of the cores and protostars.

Observations with the {\it Spitzer} Space Telescope have led to the discovery of a new class of objects called the very low luminosity objects (VeLLOs) in dense cores previously classified as ``starless'' based on the non-detection of any point source by the Infrared Astronomical Satellite (IRAS).
The VeLLOs are defined as objects having a very similar spectral energy distribution (SED) to that of a protostar, but a very faint internal luminosity of $\lesssim 0.2 \Lsun$ considering its uncertainty\citep{2004ApJS..154..396Y}. Their luminosities are found to be much fainter than those expected by the least massive protostar from the standard star formation theory by \citet{1987ARA&A..25...23S}. They may be extremely young protostars with very small central mass and mass accretion that has just begun, or normal protostars in a quiescent state between episodic accretion events during the normal mass accretion phase, or proto-brown dwarfs with very small mass accretion rates \citep{1990AJ.....99..869K,2011ApJ...743...98M,2012ApJ...747...52D,2014ApJ...781...33M,2016ARA&A..54..135H}. 

Several VeLLOs, including L1014-IRS ($L_{int} \sim 0.09 \Lsun$; \citealt{2004ApJS..154..396Y}), L1521F-IRS ($L_{int} \sim 0.06 \Lsun$; \citealt{2006ApJ...649L..37B}), IRAM 04191+1522 ($L_{int} \sim 0.08 \Lsun$; \citealt{2006ApJ...651..945D}), L328-IRS ($L_{int} \sim 0.05 \Lsun$; \citealt{2009ApJ...693.1290L}), L673-7-IRS ($L_{int} \sim 0.04 \Lsun$; \citealt{2010ApJ...721..995D}), L1148-IRS ($L_{int} \sim 0.10 \Lsun$; \citealt{2011MNRAS.416.2341K}), IC348-SMM2E ($L_{int} \sim 0.06 \Lsun$; \citealt{2014MNRAS.444..833P}), and IRAS 16253-2429 ($L_{int} \sim 0.08 \Lsun$; \citealt{2016ApJ...826...68H}), have been studied in detail. Irrespective of the basic nature of these VeLLOs, it is possible that they are formed as a result of infall motions of material inside their parent core, which might have occurred due to mainly gravity-driven kinematics. Therefore, observations to look for the infall motions in the cores that harbor VeLLOs is important to understand the initial processes that lead to the formation of protostars or proto-brown dwarfs.

The physical and kinematic states of cores with VeLLOs remain poorly understood. So far, there are only a few cases where infalling motions in dense cores with VeLLOs have been detected (e.g., L328-IRS and L1521F-IRS). \citet{2013ApJ...777...50L} detected extended infalling motions in a core with the VeLLO, L328-IRS. The infall speeds calculated using a two-layer fitting analysis of HCN 1-0 F=2-1 are found to be $0.03-0.08~\kms$. Also, the recent study of L1521F-IRS shows an infall speed of about 0.1 $\kms$ calculated from $\rm DCO^+$ lines and the HILL5 model \citep{2016ApJ...833...97K}. The infall speeds of the cores with VeLLOs seem to be similar to those of starless cores having velocities $\leq 0.1~\kms$ \citep{2001ApJS..136..703L}. 
 
Based on the shape of the SEDs from the near-IR to the submillimeter wavelengths using photometric data from the {\it Spitzer} and the {\it Herschel} telescopes, and from the detectability in a high density tracer, $\NH$ line, \citet{2016ApJS..225...26K} recently produced a catalog of VeLLOs. In this paper, we present the results of a systematic study conducted on a statistically significant sample of VeLLOs selected from the above-mentioned catalog. The cores harboring VeLLOs were observed using HCN (J = 1-0) and $\NH$ (J = 1-0) molecular lines to characterize the infall motions in the envelopes and compare them with those of the dense starless and proto-stellar cores. This study would provide useful insights to understand the fundamental physics associated with inward motions in the envelopes where VeLLOs are currently forming and discuss their properties which would in turn help us to identify their nature.

This paper is organized as follows. In Section 2, we describe our observations and data reduction procedure. Our results and discussion are presented in Sections 3 and 4, respectively. In Section 5, we conclude the paper with a summary of the results.

\section{Observations and Data Reduction}
\subsection{Observing targets and lines}
The samples of VeLLOs studied here are listed in Table 1. The targets were selected from the catalog of the VeLLOs produced by \citet{2016ApJS..225...26K}. Distances to the target sources are adopted from the recent catalogs \citep{2018ApJ...869...83Z, 2019ApJ...870...32K, 2019ApJ...879..125Z, 2020AA...633A..51Z} where the distances to the local molecular clouds associated with the Gould Belt regions were estimated employing the Gaia DR2 parallax measurements. The internal luminosities estimated based on these new, more reliable, distances for some of the sources in our sample are found to be $\gtrsim 0.2\Lsun$. Though this is not in accordance with the original definition of VeLLOs, we retained them in our sample list as they were not detected previously by the IRAS and hence not studied in depth.

We conducted survey observations of 66 out of 95 VeLLOs cataloged by \citet{2016ApJS..225...26K} in $\NH$ and HCN. We could not observe all the 95 VeLLOs due to the limited telescope time that was available to us. Thus, although we could observe only 70$\%$ of the sources from \citet{2016ApJS..225...26K} catalog, the number is still significant to conduct any statistical study on the infall motions in the VeLLOs. The molecular lines HCN (J = 1-0) and $\NH$ (J = 1-0) are used as optically thick and thin tracers, respectively, for the detection of the inward motions. The HCN 1-0 line is usually observed with a self-absorbed feature, in other words, a double-peaked or a skewed profile, toward dense cores when this line becomes optically thick. However, this seemingly absorbed feature can be formed by the presence of two slightly different velocity components along the line of sight. The optically thin tracer $\NH$ was observed to discriminate these two scenarios.

The HCN emission is easily detected in low-mass star-forming regions \citep{1998AJ....115.1111A,2007ApJ...664..928S,2013A&A...560A...3D,2015ApJ...802..126H}. Its high critical density ($n_{cr} \sim 10^6 ~cm^{-3}$ at T=10 K; \citealt{2017stfo.book.....K}) is expected to be useful to probe a dense gas envelope region. HCN has three hyperfine lines with different intensities of F(0-1) : F(2-1) : F(1-1) = 1 : 5 : 3 under optically thin conditions. These lines are expected to trace different regions depending on their optical depths. Therefore, the HCN 1-0 line is thought to be one of the best appropriate tracers to study inward motions of gaseous material around the VeLLOs, which are deeply embedded in the dense cores. The second tracer, $\NH$, is an ion-molecule less sensitive to the chemical evolution of dense cores. This line has been detected in various environments such as infrared dark clouds, starless cores, and protostellar cores \citep{1999ApJS..123..233L,2002ApJ...570L.101B,2006ApJS..166..567R,2011A&A...525A.141B,2016ApJ...830..127S}. This molecule has seven hyperfine structures that can characterize a dense inner region and estimate the systematic velocities of dense cores with better accuracy \citep{1997ApJ...489..719M}. 

In order to trace the infall motions based on the spectral asymmetry, it is very important to exactly measure even a very small velocity shift between the two tracers that are used for comparing the motions. Hence, the setting of accurate frequencies of HCN 1-0 and $\NH$ 1-0 lines for our observations was crucial. For the HCN 1-0 line, we used the Cologne Database for Molecular Spectroscopy \citep{2005JMoSt.742..215M} and for the $\NH$ 1-0 line, we used the spectroscopic information from \citet{2001ApJS..136..703L} to set its frequencies. The frequencies of the three hyperfine lines for HCN 1-0 are set to be 88630.4157 MHz for F=1-1, 88631.8473 MHz for F=2-1, and 88633.9360 MHz for F=0-1. The frequencies for the main and isolated hyperfine components for $\NH$ 1-0 are 93173.775 MHz and 93176.265 MHz, respectively.

\subsection{Observations with KVN telescopes}
The observations were carried out using one of the three single dishes of the Korean VLBI Network (KVN) 21 m radio telescopes during September 2013 and May 2016. We observed 60 sources in the HCN 1-0 line, 64 sources in the $\NH$ 1-0 line, and 60 sources in both lines. Sky emission was subtracted with a frequency switching mode by an 8 MHz throw. The spectra were taken with both right and left circular polarized feeds and used to improve their signal-to-noise ratio (SNR) by averaging the spectra.

The spectrometer is set to have a bandwidth of 32 MHz and 4096 channels at both frequencies in order to provide velocity resolutions of 0.026 $\kms$ at 88 GHz (for the HCN line) and 0.025 $\kms$ at 93 GHz (for the $\NH$ line). The pointing and the focus were checked every 3-4 hours using nearby known strong SiO maser sources and the pointing accuracy was within 4$\arcsec$ during the observations. The data were calibrated by the standard chopper wheel method and the line intensity was obtained in the $\rm T_A^*$ scale. The beam size and main beam efficiency of the KVN are 32$\arcsec$ and 0.4 at 86 GHz, respectively \citep{2011PASP..123.1398L}. The spectral data are reduced with CLASS software.

\subsection{Observations with the Mopra telescope}
We observed two southern hemisphere sources in the 3 mm band using the Mopra 22 m telescope located in Australia during May 2014. Unfortunately, these sources were not detected at observations with the HCN and $\NH$ mainly because of their low observing sensitivity level of 1 $\sigma_T$ $\sim$ 0.4 K in a main beam temperature scale at the set velocity resolution of 0.11 $\kms$. The observations were conducted by a position switching mode with a total (ON+OFF) integration time of 30 minutes with two polarized feeds. The Mopra spectrometer, MOPS, can cover 8 GHz bandwidth simultaneously in both the wide-band and zoom modes. The zoom mode with high resolution was tuned at 89.190 GHz with 16 windows of 137 MHz. The telescope pointing correction was performed by SiO maser sources every hour, giving a pointing accuracy better than 3$\arcsec$. The measured beam size and main beam efficiency of the Mopra are 38$\arcsec$ and 0.49 at 86 GHz, respectively \citep{2005PASA...22...62L}. The spectral lines were reduced by the Australia Telescope National Facility (ATNF) Spectral Analysis Package (ASAP) and CLASS software.

\section{results}
\subsection{Observing statistics}
We observed a total of 62 sources in HCN 1-0 and $\NH$ 1-0 lines. Of these, 47 and 48 sources were detected in HCN 1-0 and $\NH$ 1-0 lines, respectively. A total of 42 sources were detected in both the lines (Table \ref{tab:tbl0}). The remaining 11 sources were detected in only one line out of two lines. They were not detected either because they do not have enough gases or because our observing sensitivity was not enough. The peak temperatures and rms values of HCN 1-0 lines are found to be between 0.4 and 2.0 K and between 0.05 to 0.12 K for most sources, respectively. The $\NH$ 1-0 lines were observed in the rms noise level of 0.07 $\sim$ 0.15 K and their peak temperatures were about 0.4 $\sim$ 2.7 K for about 80\% of the sources. Here all temperature scales are given in the main beam temperature ($T_{MB}$) scale.

\subsection{Line profile shapes}
\begin{figure}[htb!]
  \centering
  \includegraphics[scale=0.8]{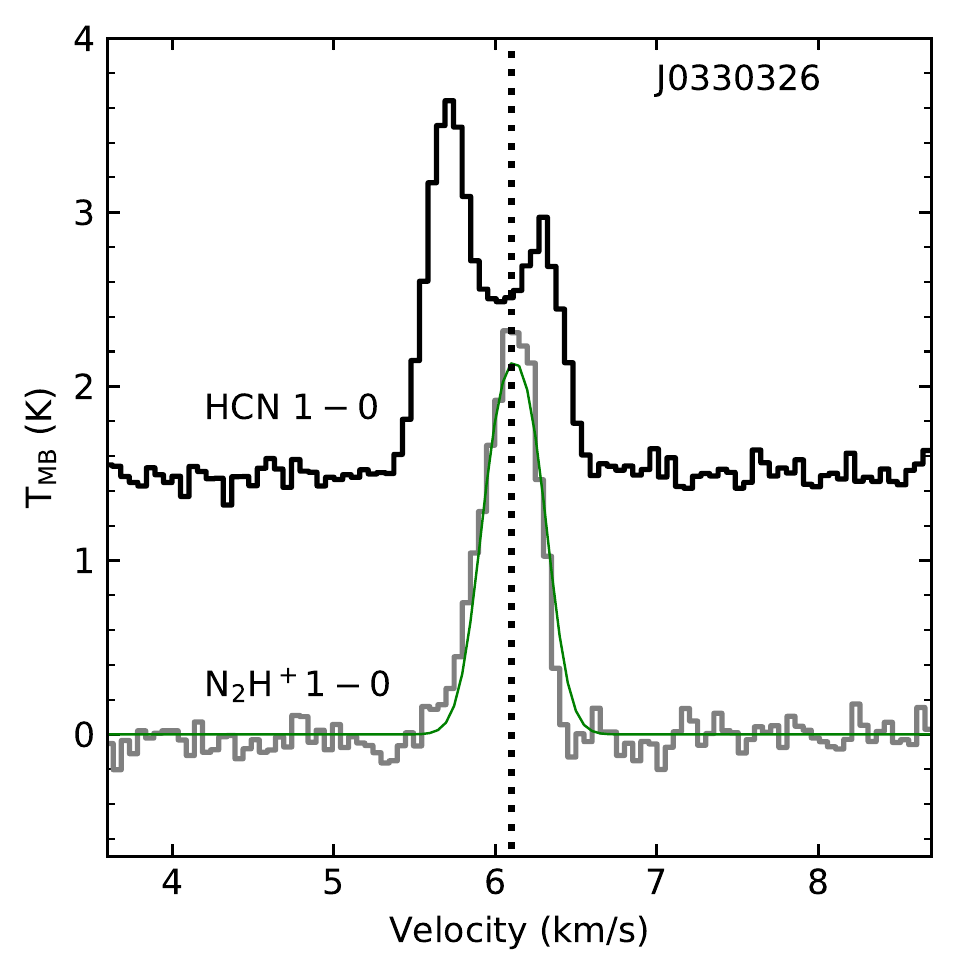}
  \caption{Observed line profiles of HCN (J=1-0 F=2-1) and $\NH$ (J=1-0 $\rm F_1F$=01-12) of J0330326. The green profile and the dotted line indicate the Gaussian profile and velocity obtained from a simultaneous Gaussian fit for $\NH$ seven hyperfine components, respectively. \label{fig:j1832424_lines}}
\end{figure}

Towards all our targets, the $\NH$ line being an optically thin tracer showed a single peaked Gaussian profile. However, HCN line profiles showed a variety of shapes; sometimes double peaked with a brighter blue peak or a brighter red peak, or single peaked with red or blue shoulder, similar to the line profiles typically observed for the starless dense cores (\citealt{2007ApJ...664..928S}). It is very interesting to note that of these various shapes seen in HCN profile, the double peaked one where the blue peak is brighter than the red peak, so called infall asymmetric profile (as shown in Figure \ref{fig:j1832424_lines}), is found to be dominant ($\sim 48\%$) compared with profiles of other shapes. The various other HCN line profiles seen in our targets are shown in Figure \ref{fig:all}.

\startlongtable
\begin{deluxetable*}{l cc cccc}
\tabletypesize{\footnotesize}
\tablewidth{0pt} 
\tablecaption{Observing targets and basic results of their observations in the HCN and $\NH$ lines \label{tab:tbl0}}
\tablehead{\colhead{Name} & \colhead{Coordinate (J2000)} & \colhead{$\rm L_{int}$} & \colhead{Distance} & \colhead{$V_{\NH}$} & \colhead{$T_{MB}^{\NH}$} & \colhead{$T_{MB}^{HCN}$} \\
           \colhead{} & \colhead{(hh:mm:ss.s dd:mm:ss)} & \colhead{$\rm (\Lsun)$} & \colhead{(pc)} & \colhead{$(\kms)$} & \colhead{$(\rm K)$} & \colhead{$(\rm K)$}}
\colnumbers
\startdata
       J0328325  &  03:28:32.5 +31:11:05  &  0.062$\pm$0.008  &  287$\pm$ 16  &   7.21$\pm$0.00  &   3.65$\pm$0.10  &   4.06$\pm$0.08 \\ 
       J0328391  &  03:28:39.1 +31:06:01  &  0.020$\pm$0.004  &  287$\pm$ 16  &   7.07$\pm$0.00  &   4.58$\pm$0.19  &   5.81$\pm$0.20 \\ 
       J0330326  &  03:30:32.6 +30:26:26  &  0.106$\pm$0.014  &  291$\pm$ 20  &   6.14$\pm$0.00  &   3.24$\pm$0.08  &   2.13$\pm$0.07 \\ 
    IC348-SMM2E  &  03:43:57.7 +32:03:10  &  1.382$\pm$0.135  &  295$\pm$ 15  &   8.70$\pm$0.00  &   2.64$\pm$0.16  &   4.11$\pm$0.10 \\ 
       J0401343  &  04:01:34.3 +41:11:43  &  0.234$\pm$0.023  &  470$\pm$ 24  &  -7.22$\pm$0.01  &   0.98$\pm$0.08  &   0.82$\pm$0.05 \\ 
IRAS04111+2800G  &  04:14:12.3 +28:08:37  &  0.069$\pm$0.007  &  141$\pm$  7  &   6.86$\pm$0.00  &   3.78$\pm$0.08  &   2.02$\pm$0.07 \\ 
       J0418402  &  04:18:40.2 +28:29:25  &  0.003$\pm$0.001  &  141$\pm$  7  &   7.27$\pm$0.01  &   1.05$\pm$0.10  &   0.61$\pm$0.08 \\ 
      IRAM04191  &  04:21:56.8 +15:29:46  &  0.047$\pm$0.004  &  141$\pm$  7  &   6.71$\pm$0.00  &   2.63$\pm$0.16  &   0.60$\pm$0.15 \\ 
     L1521F-IRS  &  04:28:38.9 +26:51:35  &  0.038$\pm$0.004  &  141$\pm$  7  &   6.49$\pm$0.00  &   3.90$\pm$0.11  &   2.93$\pm$0.08 \\ 
       J0425132  &  04:25:13.2 +26:31:45  &  0.007$\pm$0.002  &  130$\pm$  9  &         \nodata  &        $< 0.51$  &        $< 0.37$ \\ 
       J0428151  &  04:28:15.1 +36:30:28  &  0.051$\pm$0.010  &  470$\pm$ 24  &         \nodata  &        $< 0.72$  &        $< 0.32$ \\ 
       J0430149  &  04:30:14.9 +36:00:08  &  0.013$\pm$0.002  &  141$\pm$  7  &  -0.77$\pm$0.00  &   1.83$\pm$0.09  &   1.67$\pm$0.06 \\ 
       J0430559  &  04:30:55.9 +34:56:47  &  0.246$\pm$0.026  &  470$\pm$ 24  &         \nodata  &         \nodata  &   2.43$\pm$0.19 \\ 
       J0434115  &  04:34:11.5 +24:03:41  &  0.017$\pm$0.004  &  159$\pm$  8  &         \nodata  &        $< 0.51$  &   0.39$\pm$0.10 \\ 
       J0439090  &  04:39:09.0 +26:14:49  &  0.011$\pm$0.002  &  141$\pm$  7  &         \nodata  &        $< 0.96$  &        $< 0.37$ \\ 
       J0440224  &  04:40:22.4 +25:58:32  &  0.002$\pm$0.000  &  141$\pm$  7  &         \nodata  &        $< 1.84$  &        $< 0.37$ \\ 
       J1542169  &  15:42:16.9 -52:48:02  &  0.042$\pm$0.017  &  250$\pm$ 50  &         \nodata  &        $< 1.20$  &        $< 1.22$ \\ 
       J1601155  &  16:01:15.5 -41:52:35  &  0.151$\pm$0.020  &  189$\pm$ 13  &         \nodata  &        $< 1.17$  &        $< 1.21$ \\ 
       J1626484  &  16:26:48.4 -24:28:38  &  0.056$\pm$0.011  &  139$\pm$  7  &   3.65$\pm$0.00  &   0.52$\pm$0.12  &   2.79$\pm$0.18 \\ 
 IRAS16253-2429  &  16:28:21.6 -24:36:23  &  0.136$\pm$0.013  &  139$\pm$  7  &   4.08$\pm$0.00  &   5.08$\pm$0.11  &   1.16$\pm$0.21 \\ 
       J1804399  &  18:04:39.9 -04:01:22  &  0.012$\pm$0.002  &  278$\pm$ 17  &         \nodata  &        $< 0.57$  &        $< 0.44$ \\ 
       J1804493  &  18:04:49.3 -04:36:39  &  0.062$\pm$0.009  &  278$\pm$ 17  &         \nodata  &        $< 1.05$  &        $< 0.65$ \\ 
       J1809419  &  18:09:41.9 -03:31:26  &  0.020$\pm$0.004  &  278$\pm$ 17  &         \nodata  &        $< 0.80$  &         \nodata \\ 
        CB130-1  &  18:16:16.3 -02:32:37  &  0.072$\pm$0.011  &  278$\pm$ 17  &   7.60$\pm$0.00  &   2.97$\pm$0.13  &   0.68$\pm$0.10 \\ 
       L328-IRS  &  18:16:59.4 -18:02:30  &  0.172$\pm$0.046  &  217$\pm$ 30  &   6.72$\pm$0.01  &   1.01$\pm$0.11  &   1.73$\pm$0.07 \\ 
       J1828558  &  18:28:55.8 -01:37:34  &  0.516$\pm$0.193  &  484$\pm$ 96  &   7.44$\pm$0.00  &   1.29$\pm$0.16  &   1.27$\pm$0.13 \\ 
       J1829054  &  18:29:05.4 -03:42:45  &  0.602$\pm$0.225  &  484$\pm$ 96  &   5.34$\pm$0.01  &   2.10$\pm$0.21  &   0.81$\pm$0.07 \\ 
       J1829121  &  18:29:12.1 -01:48:45  &  0.028$\pm$0.011  &  484$\pm$ 96  &   6.90$\pm$0.01  &   0.57$\pm$0.10  &   1.22$\pm$0.19 \\ 
       J1829130  &  18:29:13.0 -01:46:17  &  1.261$\pm$0.470  &  484$\pm$ 96  &   6.87$\pm$0.01  &   2.81$\pm$0.28  &         \nodata \\ 
       J1829209  &  18:29:20.9 -01:37:14  &  0.653$\pm$0.244  &  484$\pm$ 96  &   7.70$\pm$0.01  &   1.77$\pm$0.26  &   1.23$\pm$0.10 \\ 
       J1829251  &  18:29:25.1 -01:47:37  &  0.215$\pm$0.082  &  484$\pm$ 96  &   7.52$\pm$0.01  &   1.20$\pm$0.14  &   1.15$\pm$0.21 \\ 
       J1829336  &  18:29:33.6 -01:45:10  &  0.083$\pm$0.032  &  484$\pm$ 96  &   7.64$\pm$0.01  &   1.50$\pm$0.09  &   4.21$\pm$0.08 \\ 
       J1829374  &  18:29:37.4 -03:14:53  &  0.381$\pm$0.143  &  484$\pm$ 96  &         \nodata  &        $< 0.81$  &        $< 0.33$ \\ 
       J1829439  &  18:29:43.9 -02:12:55  &  1.104$\pm$0.412  &  484$\pm$ 96  &   7.71$\pm$0.00  &   4.16$\pm$0.21  &         \nodata \\ 
       J1829529  &  18:29:52.9 -01:58:05  &  0.235$\pm$0.089  &  484$\pm$ 96  &   7.59$\pm$0.01  &   2.34$\pm$0.17  &   0.52$\pm$0.07 \\ 
       J1829583  &  18:29:58.3 -01:57:40  &  0.211$\pm$0.085  &  484$\pm$ 96  &   7.71$\pm$0.02  &   2.87$\pm$0.35  &        $< 0.72$ \\ 
       J1830144  &  18:30:14.4 -01:33:33  &  0.250$\pm$0.095  &  484$\pm$ 96  &   8.13$\pm$0.00  &   2.97$\pm$0.16  &   2.28$\pm$0.14 \\ 
       J1830156  &  18:30:15.6 -02:07:19  &  0.735$\pm$0.278  &  484$\pm$ 96  &   6.74$\pm$0.00  &   5.04$\pm$0.16  &   1.11$\pm$0.09 \\ 
       J1830162  &  18:30:16.2 -01:52:52  &  0.519$\pm$0.194  &  484$\pm$ 96  &   6.74$\pm$0.00  &   3.01$\pm$0.14  &   0.50$\pm$0.08 \\ 
       J1830174  &  18:30:17.4 -02:09:58  &  0.989$\pm$0.370  &  484$\pm$ 96  &   6.93$\pm$0.00  &   3.45$\pm$0.16  &   1.01$\pm$0.07 \\ 
       J1830218  &  18:30:21.8 -01:52:01  &  0.663$\pm$0.248  &  484$\pm$ 96  &   7.20$\pm$0.04  &   0.69$\pm$0.20  &   0.28$\pm$0.05 \\ 
       J1830275  &  18:30:27.5 -01:54:39  &  0.488$\pm$0.184  &  484$\pm$ 96  &   7.38$\pm$0.01  &   1.49$\pm$0.12  &   0.78$\pm$0.07 \\ 
       J1830469  &  18:30:46.9 -01:56:45  &  0.855$\pm$0.335  &  484$\pm$ 96  &   7.84$\pm$0.01  &   2.25$\pm$0.22  &         \nodata \\ 
       J1830476  &  18:30:47.6 -02:43:56  &  0.253$\pm$0.095  &  484$\pm$ 96  &         \nodata  &        $< 0.78$  &        $< 0.69$ \\ 
       J1832374  &  18:32:37.4 -02:50:45  &  0.738$\pm$0.276  &  484$\pm$ 96  &   6.24$\pm$0.01  &   0.86$\pm$0.13  &        $< 0.40$ \\ 
       J1832424  &  18:32:42.4 -02:47:56  &  0.677$\pm$0.253  &  484$\pm$ 96  &   6.30$\pm$0.00  &   1.90$\pm$0.17  &   0.85$\pm$0.08 \\ 
       J1832456  &  18:32:45.6 -02:46:57  &  0.124$\pm$0.048  &  484$\pm$ 96  &   6.33$\pm$0.01  &   0.77$\pm$0.11  &   0.56$\pm$0.14 \\ 
       J1833294  &  18:33:29.4 -02:45:58  &  0.295$\pm$0.111  &  484$\pm$ 96  &   7.36$\pm$0.00  &   2.61$\pm$0.15  &        $< 0.21$ \\ 
       J1839298  &  18:39:29.8 +00:37:40  &  0.187$\pm$0.070  &  484$\pm$ 96  &   8.22$\pm$0.01  &   0.68$\pm$0.13  &   0.44$\pm$0.14 \\ 
     L673-7-IRS  &  19:21:34.8 +11:21:23  &  0.042$\pm$0.027  &  300$\pm$100  &   7.22$\pm$0.00  &   2.15$\pm$0.09  &   2.63$\pm$0.12 \\ 
      L1148-IRS  &  20:40:56.6 +67:23:04  &  0.127$\pm$0.014  &  352$\pm$ 18  &   2.62$\pm$0.00  &   0.54$\pm$0.07  &   0.40$\pm$0.06 \\ 
       J2102212  &  21:02:21.2 +67:54:20  &  0.381$\pm$0.037  &  352$\pm$ 18  &   2.88$\pm$0.01  &   2.31$\pm$0.13  &   1.64$\pm$0.07 \\ 
       J2102273  &  21:02:27.3 +67:54:18  &  0.158$\pm$0.017  &  352$\pm$ 18  &   2.80$\pm$0.11  &   2.61$\pm$0.49  &   1.53$\pm$0.10 \\ 
      L1014-IRS  &  21:24:07.5 +49:59:08  &  0.092$\pm$0.037  &  250$\pm$ 50  &   4.20$\pm$0.01  &   0.55$\pm$0.18  &   0.69$\pm$0.13 \\ 
       J2144483  &  21:44:48.3 +47:44:59  &  0.676$\pm$0.067  &  783$\pm$ 36  &         \nodata  &         \nodata  &   0.93$\pm$0.07 \\ 
       J2144570  &  21:44:57.0 +47:41:52  &  0.066$\pm$0.015  &  783$\pm$ 36  &   1.63$\pm$0.02  &   0.57$\pm$0.12  &   0.84$\pm$0.11 \\ 
       J2145312  &  21:45:31.2 +47:36:21  &  0.525$\pm$0.052  &  783$\pm$ 36  &   3.86$\pm$0.01  &   0.96$\pm$0.10  &   0.66$\pm$0.11 \\ 
       J2146575  &  21:46:57.5 +47:32:23  &  0.510$\pm$0.063  &  774$\pm$ 41  &         \nodata  &        $< 0.96$  &   0.39$\pm$0.10 \\ 
       J2147030  &  21:47:03.0 +47:33:14  &  0.575$\pm$0.065  &  774$\pm$ 41  &   4.05$\pm$0.02  &   0.52$\pm$0.17  &   0.41$\pm$0.10 \\ 
       J2147060  &  21:47:06.0 +47:39:39  &  0.229$\pm$0.042  &  783$\pm$ 36  &   4.32$\pm$0.01  &   0.74$\pm$0.10  &   0.79$\pm$0.09 \\ 
       J2147556  &  21:47:55.6 +47:37:11  &  0.668$\pm$0.087  &  783$\pm$ 36  &         \nodata  &         \nodata  &   0.58$\pm$0.12 \\ 
       J2148585  &  21:48:58.5 +47:25:42  &  0.836$\pm$0.094  &  730$\pm$ 42  &         \nodata  &         \nodata  &        $< 0.45$ \\ 
       J2156073  &  21:56:07.3 +76:42:29  &  0.022$\pm$0.003  &  352$\pm$ 18  &         \nodata  &        $< 0.23$  &        $< 0.28$ \\ 
       J2229333  &  22:29:33.3 +75:13:16  &  0.074$\pm$0.008  &  352$\pm$ 18  &  -3.68$\pm$0.01  &   1.52$\pm$0.28  &   0.39$\pm$0.11 \\ 
       J2229594  &  22:29:59.4 +75:14:03  &  0.330$\pm$0.032  &  352$\pm$ 18  &  -4.00$\pm$0.01  &   1.43$\pm$0.14  &   0.68$\pm$0.13 \\ 
    L1251A-IRS4  &  22:31:05.5 +75:13:37  &  0.261$\pm$0.026  &  352$\pm$ 18  &  -4.19$\pm$0.00  &   3.96$\pm$0.16  &   2.48$\pm$0.09 \\ 
\enddata
\tablecomments{The source names and coordinates in Col.(1) and (2) are from \citet{2016ApJS..225...26K}. Col.(3): The internal luminosities of the sources. These were re-calculated using the new distances given in Col.(4), which are from \citet{2018ApJ...869...83Z, 2019ApJ...870...32K, 2019ApJ...879..125Z, 2020AA...633A..51Z}. Col.(5): The Gaussian fit velocities for $\NH$ lines. These were obtained from a simultaneous Gaussian fit for seven hyperfine components of the $\NH$ spectrum. Col.(6) and (7): The peak temperatures for the main components in the $\NH$ 1-0 and HCN 1-0 lines, respectively. The values for sources with no detection in the $\NH$ and HCN are given by their 3 $\sigma_{T}$ upper limit. The ellipsis symbols ($\cdots$) are meant to indicate ``no observations''.}
\end{deluxetable*}

\begin{figure*}[htb!]
  \centering
  \includegraphics[scale=0.80]{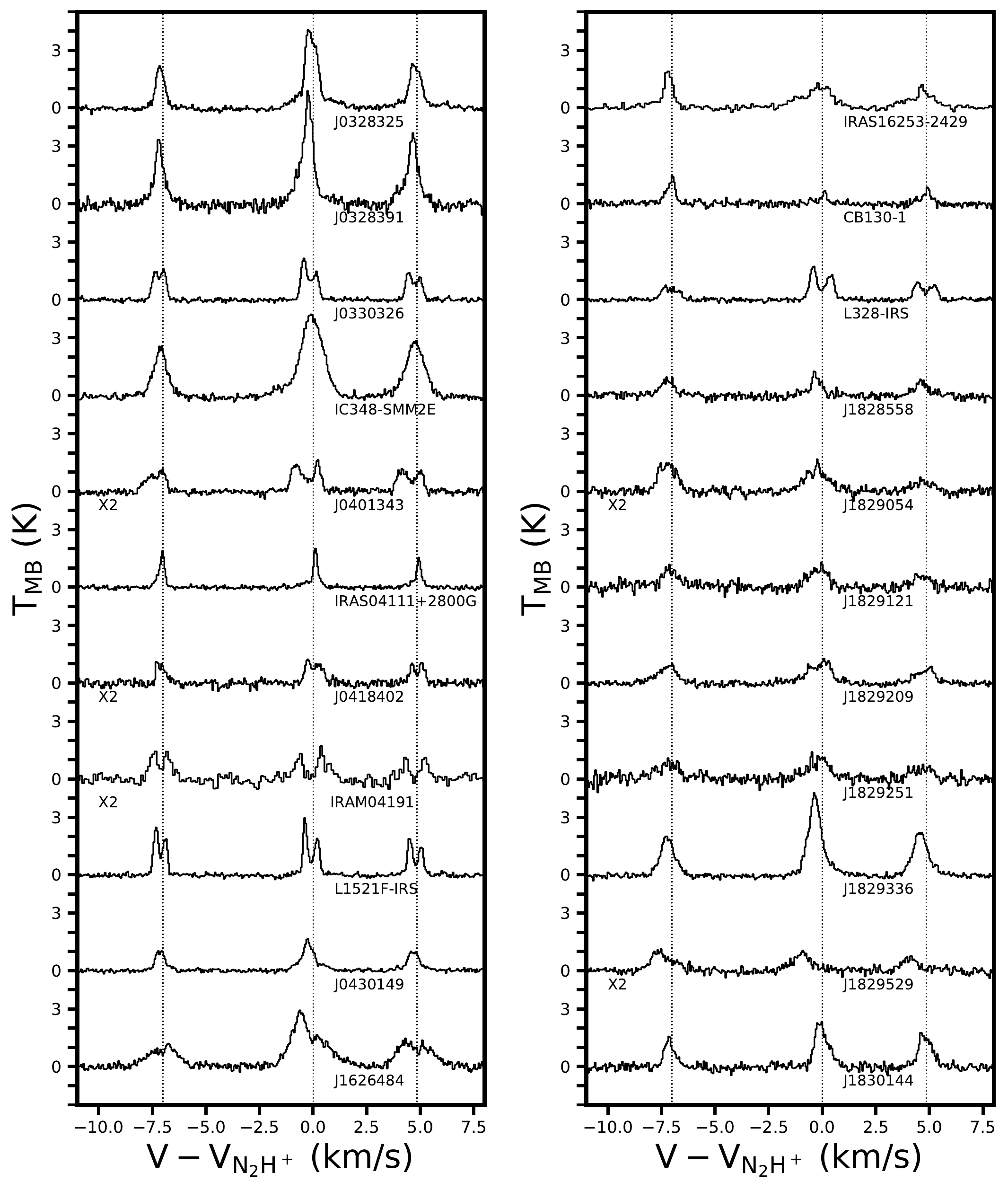}
  \caption{HCN (J=1-0) line profiles detected toward the sources. The dashed lines in the profiles indicate the $\NH$ velocities of the sources which are derived with the simultaneous Gaussian fit of the seven hyperfine components of $\NH$ line. The HCN profiles were shifted for the velocities of $\NH$ in comparison with the main component of HCN to be zero $\kms$. The spectra for IRAM04191, IRAS16253-2429, J1832456, J1839298, J2147030, and J2229333 were Hanning-smoothed. \label{fig:all}}
\end{figure*}

\setcounter{figure}{1}
\begin{figure*}[htb!]
  \centering
  \includegraphics[scale=0.83]{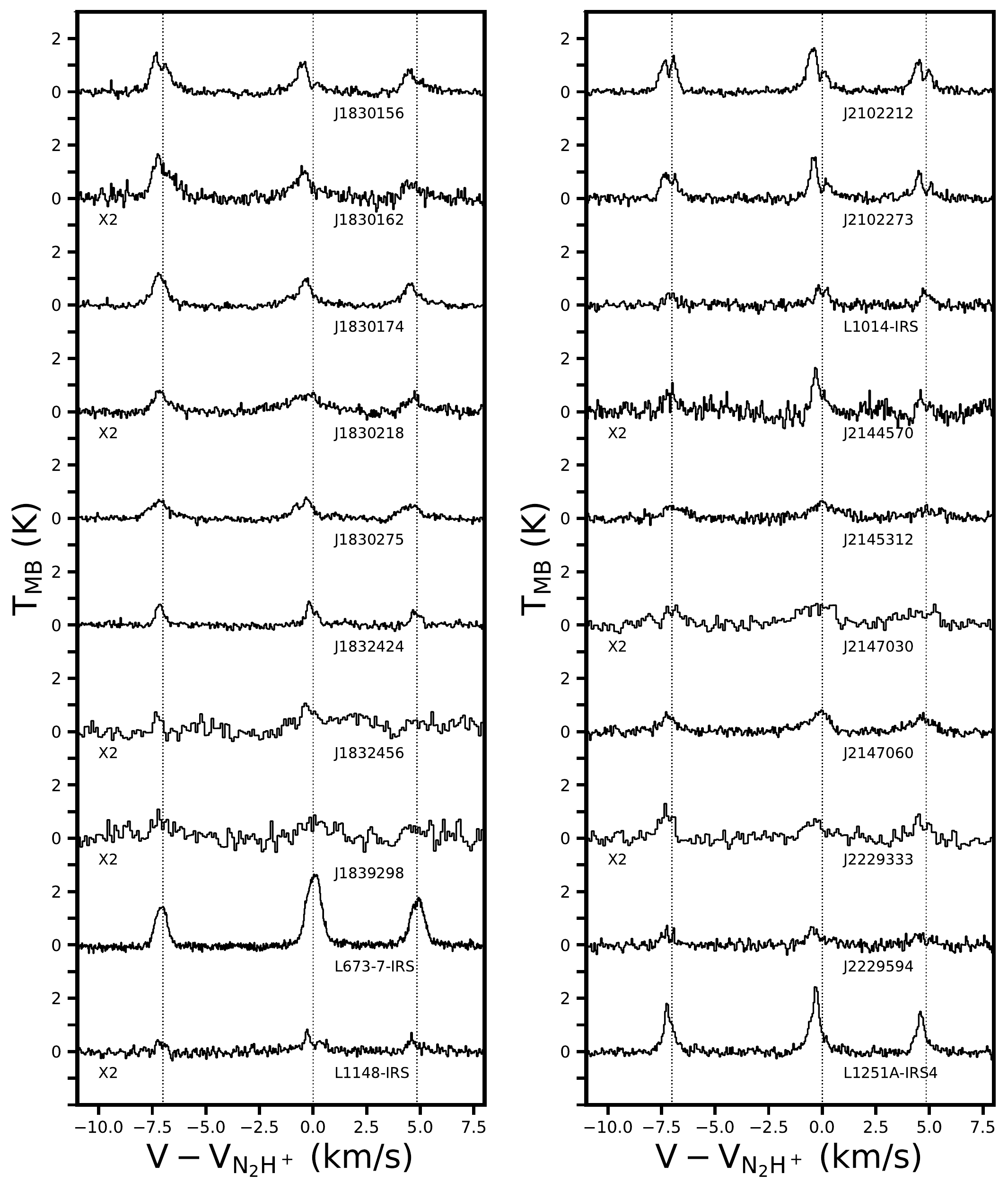}
  \caption{(Continued.)}
\end{figure*}

Based on the characteristic shapes of HCN line profiles, we classified the sources observed by us into the following categories:

\begin{verse}
{1. Sources showing blue asymmetry profiles in the main component} -- J0328325, J0328391, J0330326, IC348-SMM2E, J0418402, L1521F-IRS, J0430149, L328-IRS, J1828558, J1829054, J1830144, J1830156, J1830162, J1830174, J1832424, L1148-IRS, J2102212, J2102273, J2144570, J2229594, and L1251A-IRS4.
\end{verse}
\begin{verse}
{2. Sources showing red asymmetry profiles in the main component} -- J0401343, IRAS04111+2800G, IRAM04191, CB130-1, J1829209, and L673-7-IRS.
\end{verse}
\begin{verse}
{3. Sources showing a mixture of blue and red asymmetry in the hyperfine profiles}. This profile may be due to the presence of both expanding and infalling motions and HCN hyperfine line may be tracing these motions selectively due to their different optical depth. The sources are J0330326, J0401343, IRAM04191, J0430149, J1829121, and J2102212.
\end{verse}
\begin{verse}
{4. Sources with wide wing features}. 
We found several sources (J0328325, J0328391, IC348-SMM2E, IRAS04111+2800G, L1521F-IRS, IRAS16253-2429, J1829054, J1829336, J1830174, J2102212, J2147060, and L1251A-IRS4) that showed wide wing features in HCN profiles, probably implying the existence of outflows.
\end{verse}
\begin{verse}
{5. Sources showing anomalous intensity ratios among the hyperfine components}. In our survey, we found a number of sources that are very peculiar as they show anomalous intensity ratios among HCN hyperfine transitions. In local thermodynamic equilibrium (LTE) and optically thin conditions, the relative intensity ratios are expected to be 1:5:3 for the hyperfine lines of F=0-1:2-1:1-1, respectively. However, more than half of our samples (24) show very different ratios from the LTE values. One extreme case is J1833294, where the F=0-1 component only is seen while two other hyperfine components are completely absent. A number of factors such as turbulent overlap, radiative scattering, line opacity effects, and line overlap have been suggested as the possible cause for this observed phenomena \citep{1981A&A....97..213G,1987A&A...183L..10C,1993A&A...279..506G,1997ApJ...483..235T,2012MNRAS.420.1367L}. At this moment, any further discussion on the observed anomalous ratio is out of the scope of this work and thus we leave it for further study in the future.
\end{verse}

The number of sources in each category is counted as 21 ($\sim 48\%$) in category 1, 6 ($\sim 14\%$) in category 2 and 3, and 12 ($\sim 27\%$) in category 4. We noticed that a considerable number of sources are found to be belonging to more than one category.

\begin{deluxetable*}{lcccccc}
\tabletypesize{\footnotesize}
\tablewidth{0pt} 
\addtolength{\tabcolsep}{-2.pt}
\tablecaption{The velocity information for 41 VeLLOs by HCN and $\NH$ lines \label{tab:tbl1}}
\tablehead{\colhead{Name} & \colhead{$V_{N_2 H^+}$} & \colhead{$\Delta V_{N_2 H^+}$} & \colhead{$V_{HCN}$} & \colhead{$\delta V^{HCN}_{F=0-1}$} & \colhead{$\delta V^{HCN}_{F=2-1}$} & \colhead{$\delta V^{HCN}_{F=1-1}$} \\
           \colhead{} & \colhead{$(\rm \kms)$} & \colhead{$(\rm \kms)$} & \colhead{$(\rm \kms)$} & \colhead{} & \colhead{} & \colhead{} }
\colnumbers
\startdata
       J0328325  &   7.21$\pm$0.01  &   0.37$\pm$ 0.01  &   7.04$\pm$ 0.01  &  -0.16$\pm$ 0.01  &  -0.45$\pm$ 0.01  &  -0.48$\pm$ 0.01 \\ 
       J0328391  &   7.07$\pm$0.01  &   0.41$\pm$ 0.01  &   6.88$\pm$ 0.01  &  -0.23$\pm$ 0.01  &  -0.46$\pm$ 0.01  &  -0.37$\pm$ 0.01 \\ 
       J0330326  &   6.14$\pm$0.01  &   0.35$\pm$ 0.01  &   5.75$\pm$ 0.01  &   0.30$\pm$ 0.01  &  -1.11$\pm$ 0.01  &  -0.98$\pm$ 0.01 \\ 
    IC348-SMM2E  &   8.70$\pm$0.01  &   0.63$\pm$ 0.01  &   8.61$\pm$ 0.03  &  -0.06$\pm$ 0.01  &  -0.14$\pm$ 0.01  &  -0.16$\pm$ 0.01 \\ 
       J0401343  &  -7.22$\pm$0.01  &   0.35$\pm$ 0.01  &  -6.99$\pm$ 0.01  &   0.03$\pm$ 0.01  &   0.65$\pm$ 0.03  &  -1.83$\pm$ 0.07 \\ 
IRAS04111+2800G  &   6.86$\pm$0.01  &   0.27$\pm$ 0.01  &   7.00$\pm$ 0.01  &   0.24$\pm$ 0.01  &   0.53$\pm$ 0.01  &   0.46$\pm$ 0.01 \\ 
       J0418402  &   7.27$\pm$0.01  &   0.34$\pm$ 0.01  &   7.07$\pm$ 0.01  &  -0.49$\pm$ 0.02  &  -0.59$\pm$ 0.02  &   0.68$\pm$ 0.03 \\ 
      IRAM04191  &   6.71$\pm$0.01  &   0.51$\pm$ 0.01  &   7.14$\pm$ 0.01  &  -0.59$\pm$ 0.01  &   0.84$\pm$ 0.02  &   0.78$\pm$ 0.02 \\ 
     L1521F-IRS  &   6.49$\pm$0.01  &   0.30$\pm$ 0.01  &   6.15$\pm$ 0.01  &  -0.76$\pm$ 0.01  &  -1.12$\pm$ 0.01  &  -0.96$\pm$ 0.01 \\ 
       J0430149  &  -0.77$\pm$0.01  &   0.31$\pm$ 0.01  &  -1.01$\pm$ 0.01  &   0.08$\pm$ 0.01  &  -0.77$\pm$ 0.02  &  -0.73$\pm$ 0.02 \\ 
 IRAS16253-2429  &   4.08$\pm$0.01  &   0.24$\pm$ 0.01  &   4.06$\pm$ 0.04  &  -0.40$\pm$ 0.01  &  -0.09$\pm$ 0.01  &  -0.52$\pm$ 0.01 \\ 
        CB130-1  &   7.60$\pm$0.01  &   0.34$\pm$ 0.01  &   7.74$\pm$ 0.01  &   0.33$\pm$ 0.01  &   0.42$\pm$ 0.01  &   0.36$\pm$ 0.01 \\ 
       L328-IRS  &   6.72$\pm$0.01  &   0.61$\pm$ 0.03  &   6.32$\pm$ 0.01  &  -0.38$\pm$ 0.02  &  -0.66$\pm$ 0.03  &  -0.61$\pm$ 0.03 \\ 
       J1828558  &   7.44$\pm$0.01  &   0.64$\pm$ 0.03  &   7.11$\pm$ 0.03  &  -0.28$\pm$ 0.01  &  -0.51$\pm$ 0.03  &  -0.35$\pm$ 0.02 \\ 
       J1829054  &   5.34$\pm$0.01  &   0.64$\pm$ 0.03  &   5.13$\pm$ 0.01  &  -0.17$\pm$ 0.01  &  -0.32$\pm$ 0.02  &  -0.26$\pm$ 0.01 \\ 
       J1829121  &   6.90$\pm$0.01  &   0.46$\pm$ 0.04  &   6.89$\pm$ 0.02  &  -0.16$\pm$ 0.01  &  -0.01$\pm$ 0.01  &  -0.85$\pm$ 0.07 \\ 
       J1829209  &   7.70$\pm$0.01  &   0.49$\pm$ 0.03  &   7.86$\pm$ 0.02  &   0.05$\pm$ 0.01  &   0.32$\pm$ 0.02  &   0.47$\pm$ 0.03 \\ 
       J1829251  &   7.52$\pm$0.01  &   0.44$\pm$ 0.02  &   7.54$\pm$ 0.03  &  -0.17$\pm$ 0.01  &   0.04$\pm$ 0.01  &  -0.26$\pm$ 0.01 \\ 
       J1829336  &   7.64$\pm$0.01  &   0.63$\pm$ 0.01  &   7.32$\pm$ 0.01  &  -0.32$\pm$ 0.01  &  -0.51$\pm$ 0.01  &  -0.41$\pm$ 0.01 \\ 
       J1829529  &   7.59$\pm$0.01  &   0.62$\pm$ 0.02  &   6.72$\pm$ 0.02  &  -0.80$\pm$ 0.02  &  -1.41$\pm$ 0.04  &  -1.09$\pm$ 0.03 \\ 
       J1830144  &   8.13$\pm$0.01  &   0.34$\pm$ 0.01  &   8.05$\pm$ 0.01  &  -0.21$\pm$ 0.01  &  -0.24$\pm$ 0.01  &  -0.50$\pm$ 0.01 \\ 
       J1830156  &   6.74$\pm$0.01  &   0.53$\pm$ 0.01  &   6.27$\pm$ 0.01  &  -0.48$\pm$ 0.01  &  -0.90$\pm$ 0.01  &  -0.62$\pm$ 0.01 \\ 
       J1830162  &   6.74$\pm$0.01  &   0.51$\pm$ 0.01  &   6.32$\pm$ 0.02  &  -0.26$\pm$ 0.01  &  -0.83$\pm$ 0.02  &  -0.90$\pm$ 0.02 \\ 
       J1830174  &   6.93$\pm$0.01  &   0.50$\pm$ 0.01  &   6.62$\pm$ 0.01  &  -0.20$\pm$ 0.01  &  -0.62$\pm$ 0.01  &  -0.56$\pm$ 0.01 \\ 
       J1830218  &   7.20$\pm$0.04  &   0.97$\pm$ 0.13  &   7.06$\pm$ 0.03  &  -0.14$\pm$ 0.02  &  -0.14$\pm$ 0.02  &  -0.13$\pm$ 0.02 \\ 
       J1830275  &   7.38$\pm$0.01  &   0.76$\pm$ 0.02  &   7.12$\pm$ 0.01  &  -0.11$\pm$ 0.01  &  -0.34$\pm$ 0.01  &  -0.10$\pm$ 0.01 \\ 
       J1832424  &   6.30$\pm$0.01  &   0.29$\pm$ 0.01  &   6.16$\pm$ 0.02  &  -0.24$\pm$ 0.01  &  -0.47$\pm$ 0.02  &  -0.43$\pm$ 0.02 \\ 
       J1832456  &   6.33$\pm$0.01  &   0.37$\pm$ 0.02  &   6.22$\pm$ 0.04  &          \nodata  &  -0.30$\pm$ 0.02  &          \nodata \\ 
       J1839298  &   8.22$\pm$0.01  &   0.40$\pm$ 0.03  &   8.30$\pm$ 0.10  &          \nodata  &   0.19$\pm$ 0.02  &          \nodata \\ 
     L673-7-IRS  &   7.22$\pm$0.01  &   0.56$\pm$ 0.01  &   7.39$\pm$ 0.01  &   0.05$\pm$ 0.01  &   0.30$\pm$ 0.01  &   0.20$\pm$ 0.01 \\ 
      L1148-IRS  &   2.62$\pm$0.01  &   0.21$\pm$ 0.01  &   2.38$\pm$ 0.01  &  -0.73$\pm$ 0.04  &  -1.14$\pm$ 0.06  &  -1.05$\pm$ 0.06 \\ 
       J2102212  &   2.88$\pm$0.01  &   0.69$\pm$ 0.02  &   2.46$\pm$ 0.01  &   0.22$\pm$ 0.01  &  -0.62$\pm$ 0.02  &  -0.51$\pm$ 0.01 \\ 
       J2102273  &   2.80$\pm$0.11  &   0.60$\pm$ 0.07  &   2.42$\pm$ 0.01  &  -0.39$\pm$ 0.05  &  -0.63$\pm$ 0.08  &  -0.52$\pm$ 0.07 \\ 
      L1014-IRS  &   4.20$\pm$0.01  &   0.25$\pm$ 0.01  &   4.06$\pm$ 0.02  &  -0.35$\pm$ 0.02  &  -0.56$\pm$ 0.03  &  -0.26$\pm$ 0.01 \\ 
       J2144570  &   1.63$\pm$0.02  &   0.47$\pm$ 0.04  &   1.36$\pm$ 0.01  &   0.27$\pm$ 0.02  &  -0.58$\pm$ 0.05  &  -0.47$\pm$ 0.04 \\ 
       J2145312  &   3.86$\pm$0.01  &   0.62$\pm$ 0.03  &   3.88$\pm$ 0.02  &  -0.01$\pm$ 0.01  &   0.04$\pm$ 0.01  &   0.06$\pm$ 0.01 \\ 
       J2147030  &   4.05$\pm$0.02  &   0.44$\pm$ 0.04  &   3.98$\pm$ 0.06  &   0.65$\pm$ 0.05  &  -0.16$\pm$ 0.01  &   0.97$\pm$ 0.08 \\ 
       J2147060  &   4.32$\pm$0.01  &   0.62$\pm$ 0.03  &   4.27$\pm$ 0.02  &  -0.29$\pm$ 0.01  &  -0.08$\pm$ 0.01  &  -0.32$\pm$ 0.01 \\ 
       J2229333  &  -3.68$\pm$0.01  &   0.35$\pm$ 0.03  &  -3.95$\pm$ 0.07  &  -0.63$\pm$ 0.05  &  -0.77$\pm$ 0.06  &  -0.94$\pm$ 0.08 \\ 
       J2229594  &  -4.00$\pm$0.01  &   0.49$\pm$ 0.01  &  -4.41$\pm$ 0.02  &  -0.34$\pm$ 0.01  &  -0.84$\pm$ 0.02  &  -0.57$\pm$ 0.02 \\ 
    L1251A-IRS4  &  -4.19$\pm$0.01  &   0.35$\pm$ 0.01  &  -4.45$\pm$ 0.01  &  -0.43$\pm$ 0.01  &  -0.74$\pm$ 0.01  &  -0.61$\pm$ 0.01 \\ 
\enddata
\tablecomments{Col.(1) lists the source names. Col.(2) and (3) indicate the Gaussian fit velocities and FWHMs of $\NH$ (1-0). The intensity-peak velocity of HCN (1-0) line given in Col.(4) was obtained by a Gaussian fit for its brighter peak component after masking the other fainter component. The normalized velocity differences in Cols.(5)-(7) are calculated using equation (1) in the text. The uncertainties are propagation errors in $\delta V$. Line information is measured from the spectra with a channel width of 15.63 kHz (0.05 $\kms$). The ellipsis symbols ({$\cdots$}) are to indicate no-determination of $\delV$ due to the noisy profile.}
\end{deluxetable*}

\section{Discussion}
\subsection{$\delta V$ distribution of HCN spectra}
As shown in Figure \ref{fig:all}, the HCN line profiles can have various shapes. This can be quantitatively discussed by introducing the velocity differences ($\delta V$) between two intensity peaks in optically thick and thin tracers. We estimated the $\delV$ from our observed HCN 1-0 (optically thick) and $\NH$ 1-0 (optically thin) lines using the equation \citep{1997ApJ...489..719M}:
\begin{equation}
	\delta V_{\rm {HCN}} = \frac{V_{\rm {HCN}} - V_{\NH}}{\Delta V_{\NH}},
\end{equation}
where $V_{\rm {HCN}}$ is the peak velocity of the HCN line profile, $V_{\NH}$ is the Gaussian fit velocity of the $\NH$ line profile, and $\Delta V_{\NH}$ is the FWHM of the $\NH$ spectrum. The $V_{\rm {HCN}}$ was obtained for each hyperfine line of HCN by creating a Gaussian fit to the brightest peak part of the asymmetric line profile. The $V_{\NH}$ and $\Delta V_{\NH}$ were derived by making a simultaneous Gaussian fit to the seven hyperfine components of the $\NH$ line whose line parameters are given by \citet{1995ApJ...455L..77C}\footnote[1]{J1626484 has an unusually wide line width, as shown in Figure \ref{fig:all}. \citet{2005ApJ...630..381B} reported that J1626484 consists of multiple cores from the mid-infrared observation. We have confirmed that the three cores are distributed within the KVN beam size in the {\it Spitzer} 8 $\micron$ image, which can not be resolved by our HCN observations. We exclude J1626484 in further $\delta V$ discussions, even though J1626484 has a distinct blue asymmetric line profile.}.

The shapes of the HCN line profiles can be due to various factors such as optical depth, infall/expanding motions, and outflow motions. Therefore, understanding the distribution of the normalized velocity difference ($\delta V$) would be useful to recognize the dominant gas motion present around the envelopes in a quantitative way and compare them with those for the starless and other protostellar cores.

\begin{figure*}[htb!]
\centering
\includegraphics[scale=0.8]{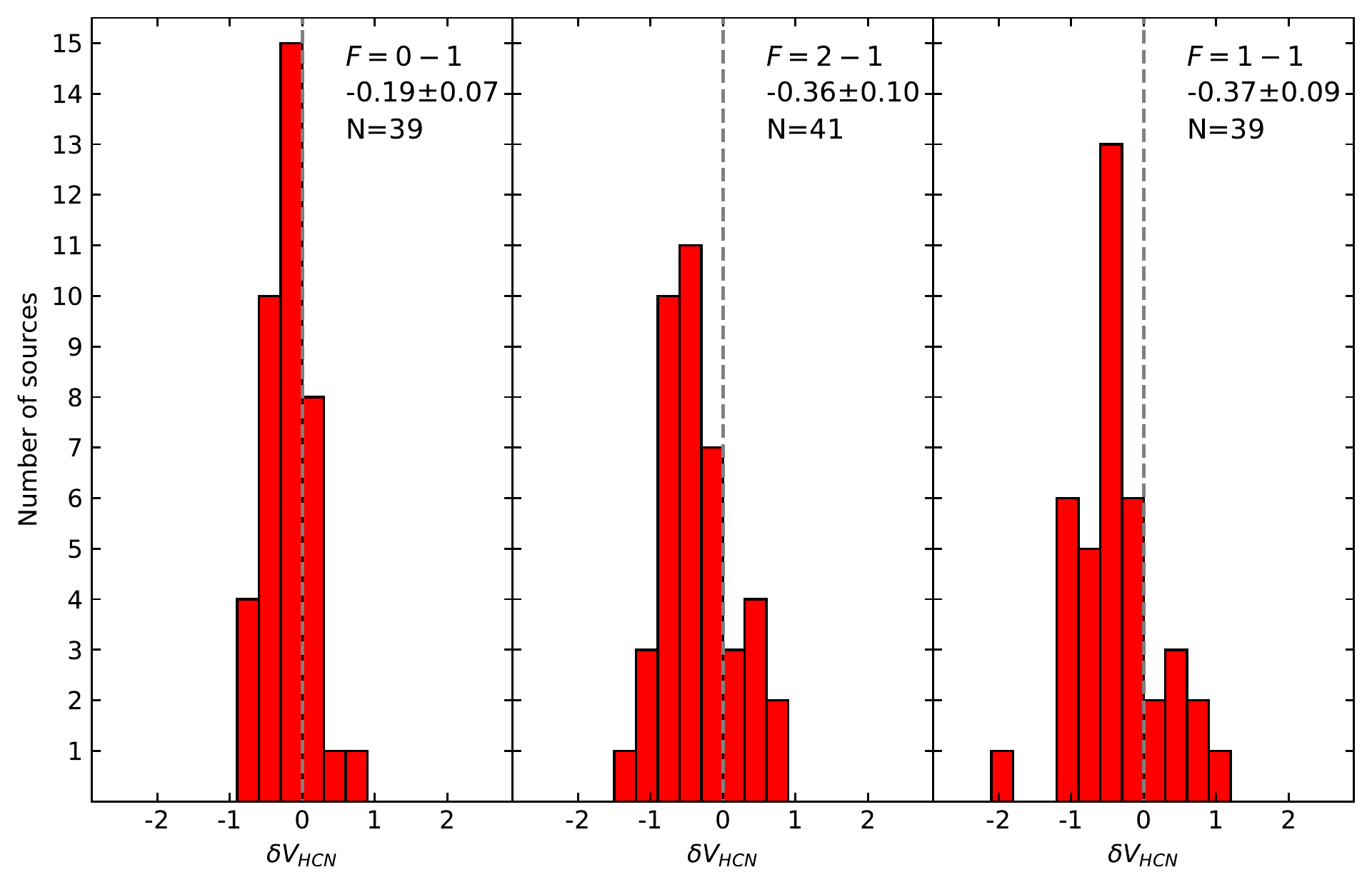}
\caption{Histograms of the normalized velocity difference between HCN (1-0) three hyperfine lines and $\NH$ (1-0) line for cores harboring VeLLOs. The legend in the right-top corners in each panel indicates a transition of three hyperfine lines of HCN (1-0), the mean $\delV$ for each hyperfine component and the standard error of the mean (SEM), and the number of sources considered. \label{fig:delV}}
\end{figure*}

We estimated the values of $\delta V$ using the three hyperfine components of HCN. The values are listed in Table \ref{tab:tbl1} and displayed as histograms in Figure \ref{fig:delV}. The distribution shown in the figure suggests that the hyperfine component with the higher optical depth has a more skewed $\delta V$ distribution towards the blue side. Among the three hyperfine components, the $\delta V$ values estimated for the main component of HCN (HCN 1-0, F=2-1) shows the largest negative shift implying that the main component may be the best tracer of inward motions in the envelope of the VeLLOs as was found true for the $\delta V$ distribution of the starless dense cores also \citep{2007ApJ...664..928S}.

\begin{figure*}[htb!]
  \centering
  \includegraphics[scale=0.8]{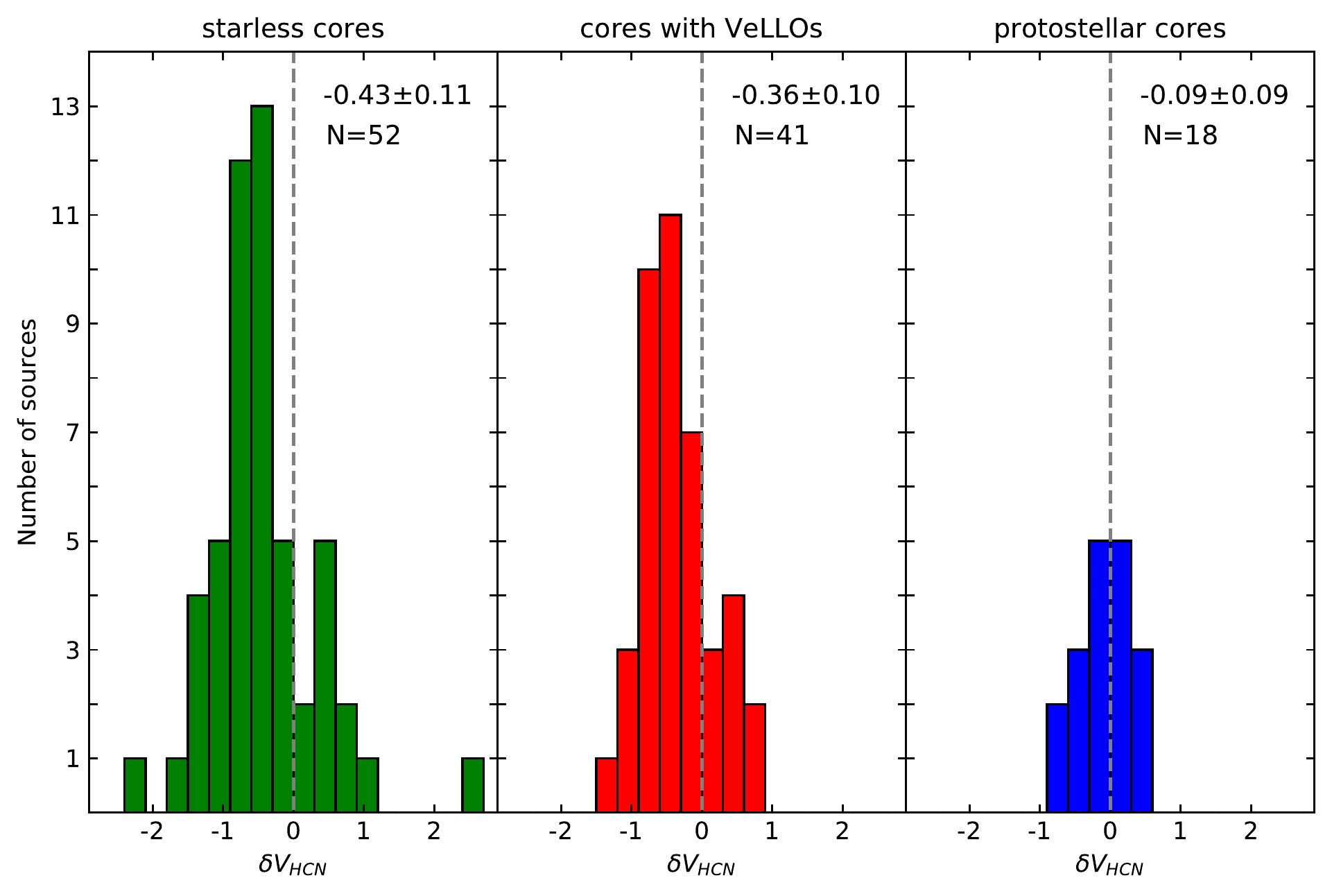}
  \caption{Histograms of the normalized velocity difference for the starless cores (green), cores with VeLLOs (red), and protostellar cores (blue). The protostellar and starless cores data were given from \citet{1999ApJ...520..223P} and \citet{2007ApJ...664..928S}, respectively. The values in the right-top corners in each panel indicate the mean and the SEM of each group and the number of sources considered. \label{fig:number_type}}
\end{figure*}

Using the HCN main component and $\NH$ lines, thus, we calculated the $\delta V$ values for the 41 cores with VeLLOs and compared them with those for the protostars \citep{1999ApJ...520..223P} and the starless cores \citep{2007ApJ...664..928S} as shown in Figure \ref{fig:number_type}. The targets in the two surveys used here for comparison were selected to include as many sources as possible which are observable in the northern sky using the TRAO 14-m telescope to avoid any likely selection bias. Note that most of the targets in these surveys are as distant as our targets and thus our KVN survey traces a slightly ($\sim 56$ percent) smaller size scale than that by TRAO telescope for the other two surveys. The results in the figure indicate that the $\delta V$ distribution of the cores with VeLLOs is more similar to that of the starless cores than that of the protostellar cores. The mean values of the $\delta V$ distributions for the starless cores, the cores with VeLLOs, and the protostellar cores are $-0.43$, $-0.36$, and $-0.09$, respectively. This may indicate that the VeLLOs are deeply embedded in gaseous envelopes, which are dominated by infalling motions, while other kinematic effects such as outflow are minimal in comparison to the normal protostars whose envelopes are highly affected by the outflow motions as well as the infalling motions.

\subsection{Infall Kinematics toward VeLLOs}
To investigate the kinematics of the infall motion in the envelopes that harbor the VeLLOs, we have calculated the full width at half maximum (FWHM) for thermal and non-thermal motions by using $\NH$ line profiles for the sources showing the infall asymmetry, $\Delta V_{T} = \sqrt{8 {\rm ln(2)} k_B T_k/\left<m \right>}$ and $\Delta V_{NT}$ = $\sqrt{\Delta V_{\NH}^2 - \left( 8 {\rm ln(2)} k_B T_k/\left<m_{\NH} \right> \right)}$, where $k_B$ is the Boltzmann constant, $T_k$ is the gas kinetic temperature, $\left<m \right>$ is the mean molecular weight (2.3 amu), and $\left<m_{\NH} \right>$ is the molecular weight of $\NH$ (29 amu). If the dust and gas are well-mixed, then the gas kinetic temperature would follow the dust temperature. With this assumption, we used the dust temperatures in the calculation of the thermal line width which are obtained from the {\it Herschel} Gould Belt survey archive\footnote{\url{http://www.herschel.fr/cea/gouldbelt/en/Phocea/Vie_des_labos/Ast/ast_visu.php?id_ast=66}}. The dust temperatures in the archive were given for each pixel which size is about 3$\arcsec$ and thus the dust temperature for each source was re-calculated as the mean value within our observing beam size of 32$\arcsec$. The temperature was found to range between 11 and 16 K with the uncertainty of $\sim 5\%$ in its value. This gives the thermal line widths (FWHMs) of 0.47 - 0.58 $\kms$ for our targets with their propagation uncertainty of $\sim 0.01~\kms$.

\begin{figure*}[htb!]
  \centering
  \includegraphics[scale=0.8]{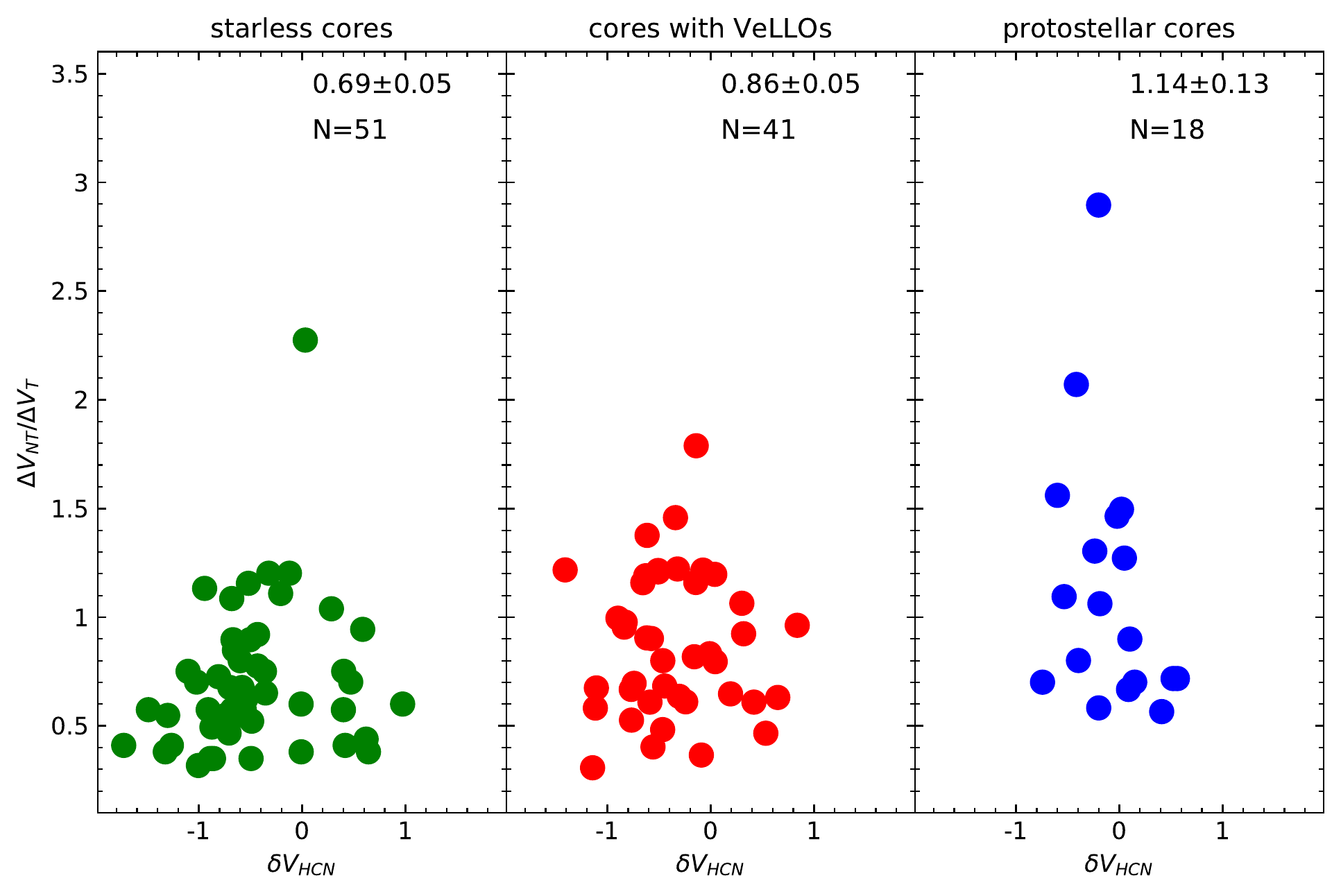}
  \caption{The distribution of the ratios of thermal and non-thermal motion components as a function of the normalized velocity differences for the starless cores (green), cores with VeLLOs (red), and protostellar cores (blue). The numeric values in the right-top corners in each panel indicate the mean and the SEM for the ratios between two motion components of each group and the number of sources considered. \label{fig:delv_motion}}
\end{figure*}

Figure \ref{fig:delv_motion} illustrates the distribution of the ratios of thermal and non-thermal motion components as a function of the normalized velocity differences for HCN main hyperfine component for the starless cores, cores with VeLLOs, and protostellar cores. For the majority (85\%) of the starless cores, this ratio is less than 1, indicating that most of the starless cores are in a state of thermal infall \citep{1999ApJS..123..233L,2007ApJ...664..928S}, while the ratio in half ($\sim 55 \%$) of the protostellar cores is greater than 1 \citep{1997ApJ...489..719M,1999ApJ...520..223P}, indicating that in some of the protostellar cores the non-thermal rather than the thermal infalling motions are dominant. In the case of cores with VeLLOs, the distribution of $\Delta V_{NT} / \Delta V_T$ is similar to or slightly higher than that of the starless cores. Protostellar cores appear to have a large influence on the non-thermal components. However, the non-thermal effect in the cores with VeLLOs is weaker, being similar to that found in the starless cores. This is probably because although they have some star-forming activities such as the presence of outflows, the energetics involved may be smaller when compared to those in the protostellar cores. 

\subsection{Infall speeds in the envelopes of the VeLLOs and their implication}
We derived the infall speed for the cores with VeLLOs showing infall signatures using the HILL5 model \citep{2005ApJ...620..800D}. The HILL5 is a radiative transfer model in which two layers are approaching each other with a speed $V_{infall}$. The excitation temperature in the layers is assumed to be peaked at a temperature $T_p$ at the center of the layer while it decreases at their near and far edges. In other words, the excitation temperature profile has a $\Lambda$-shaped distribution to the optical depths (i.e., $T_{ex}$ increases linearly with $\tau_r$ from edge to center in the rear section of the cloud and $T_{ex}$ decreases with $\tau_f$ from the center to edge in the front section of the cloud). The brightness temperature of the model is given by: 
\begin{eqnarray}
	\Delta T_B & = & [J(T_p) - J(T_0)]\bigg[\frac{(1-e^{-\tau_f})}{\tau_f-e^{-\tau_f}(1-e^{-\tau_r})/\tau_r}\bigg]\nonumber \\ 
	& & + [J(T_0) - J(T_b)][1-e^{-\tau_f - \tau_r}],
\end{eqnarray}
where $J(T) \equiv T_0/{(\exp(T_0/T)-1)}$, $T_0 = h\nu/k_B$, $h$ is Planck's constant, $\nu$ is the frequency of the transition, $k_B$ is Boltzmann's constant, and $T_b$ is the background temperature, respectively. The optical depths are given by $\tau_f = \tau_0 \exp[-(v-v_{sys}-V_{infall})^2/2\sigma_v^2]$ in the front section and $\tau_r = \tau_0 \exp[-(v-v_{sys}+V_{infall})^2/2\sigma_v^2]$ in the rear section, where $V_{infall}$ is the infall speed, $\sigma_v$ is the velocity dispersion, and $\tau_0$ is the total optical depth. The above equation is composed of five free parameters namely $T_p$, $\tau_{0}$, $V_{infall}$, $v_{sys}$, and $\sigma_v$ \citep{2005ApJ...620..800D}.

\begin{figure}[htb!]
  \centering
  \includegraphics[scale=0.8]{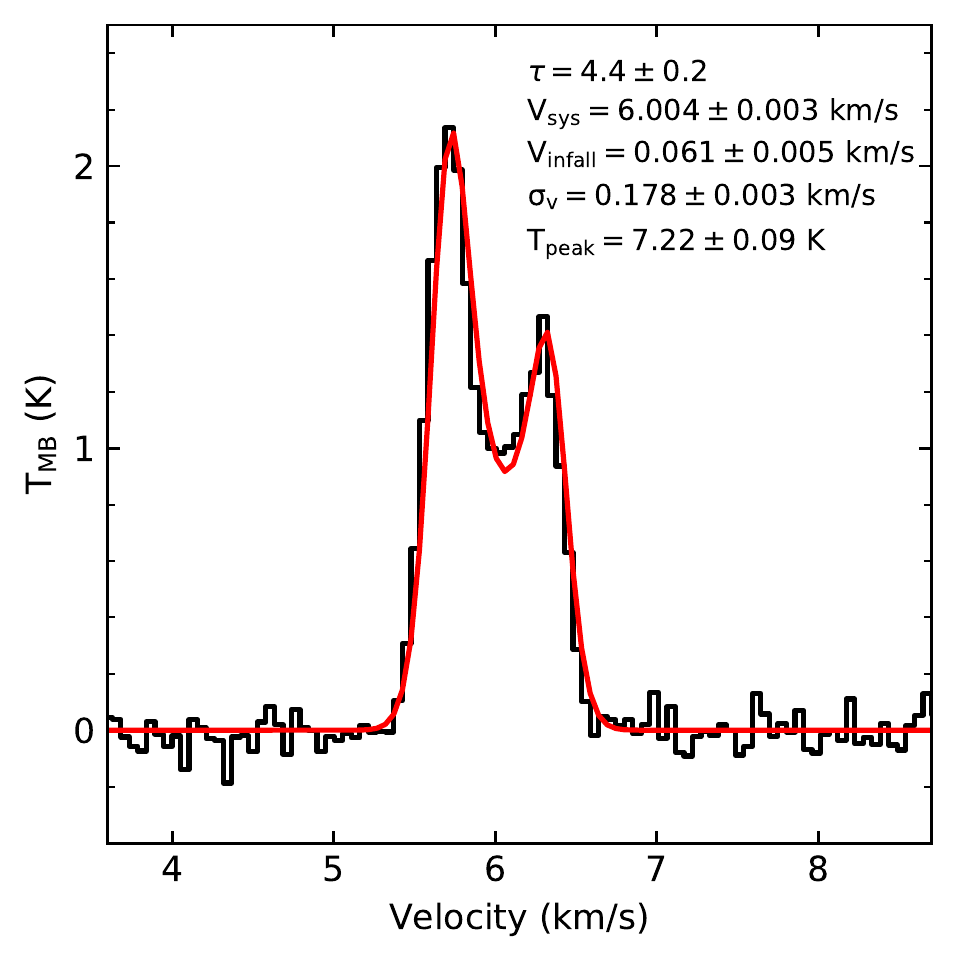}
  \caption{Best-fit results of the HILL5 model in HCN (J=1-0 F=2-1) spectrum of J0330326. The overlaid red profile shows a best-fit one of the HILL5 model. The errors for the fit parameters given in the legend are 1 $\sigma$ uncertainties of the parameters obtained from the HILL5 model fits for 1000 artificial spectra produced by the best fit parameters and observational rms. \label{fig:model_fit}}
\end{figure}

We fitted 15 sources for which the HCN 1-0 spectra show clear double peaks or red shoulder, using HILL5 via the PySpecKit python module\footnote{\url{http://pyspeckit.bitbucket.org}} \citep{Ginsburg:2011vg}. The fits require an input guess as a starting point for the fitting algorithm. The guesses for $v_{sys}$ and $\sigma_v$ were set based on the observational values obtained from $\NH$ lines. The best fit parameter for $v_{sys}$ was obtained within the range of $\pm$0.1 $\kms$ from the observational value of $V_{\NH}$ while $\sigma_v$ was fixed by the line width of $\NH$. We performed HILL5 model fits for 1000 spectra, which are simulated with the best fit parameters, with random noise equivalent to the observing rms for every spectrum to calculate 1 $\sigma$ uncertainties on the parameters. The result of HILL5 fitting for the HCN (J=1-0 F=2-1) spectrum of J0330326 as a fitting example is shown in Figure \ref{fig:model_fit}, indicating that the asymmetric shape of the observed HCN (J=1-0 F=2-1) line is well fitted with the HILL5 model within a precision of 1 $\sigma=0.05~\kms$.

\begin{figure}[htb!]
  \centering
  \includegraphics[scale=0.8]{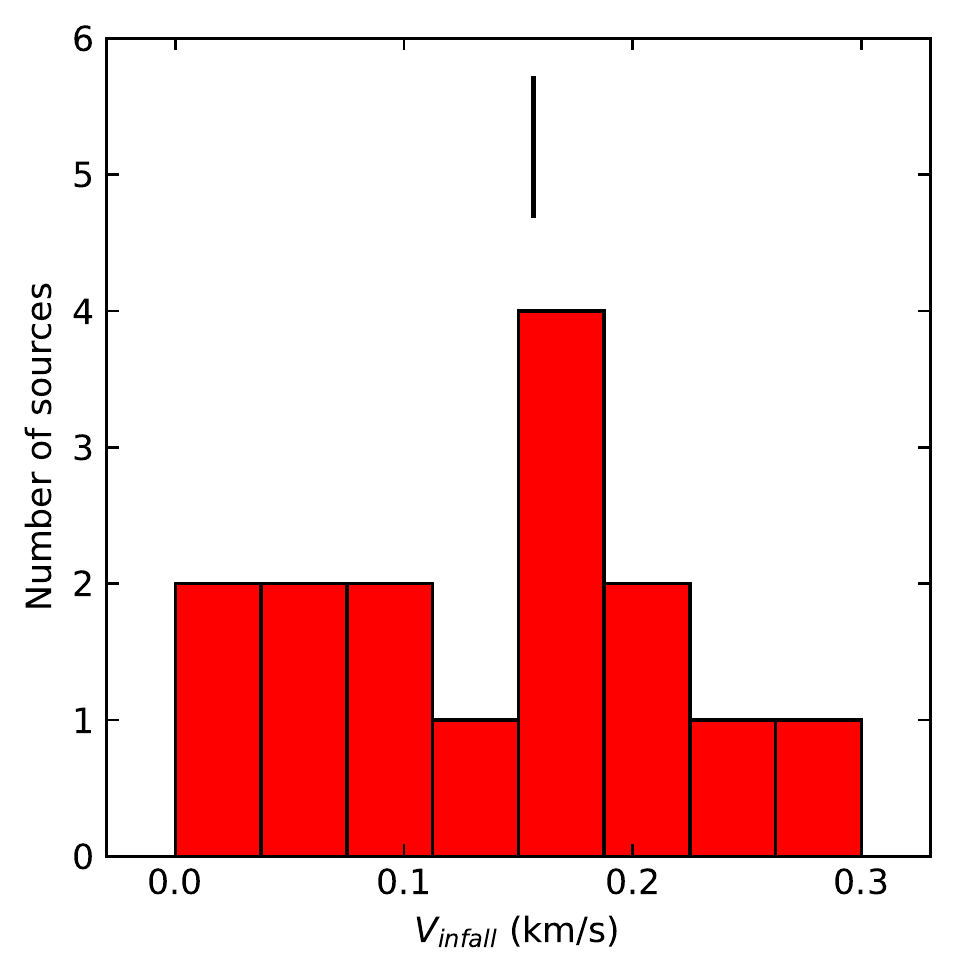}
  \caption{Histogram of the infall speeds for 15 cores with VeLLOs. The vertical line indicates the median value of 0.16 $\kms$. \label{fig:Vinfall}}
\end{figure}

The infall speeds of the 15 VeLLOs obtained from the HILL5 model fits are listed in Table \ref{tab:tbl3}. Figure \ref{fig:Vinfall} displays the distribution of these infall speeds, showing that the derived infall speed ranges mostly from 0.03 $\kms$ to 0.3 $\kms$, with a median value of 0.16 $\kms$. 
The infall speeds derived using similar models towards the starless cores and protostellar cores are found to be $V_{infall} \sim 0.05-0.09~\kms$ \citep{2001ApJS..136..703L} and $V_{infall} \sim 0.3-0.7~\kms$ \citep{2001ApJ...562..770D,2013A&A...558A.126M}, respectively, suggesting that the infall speed of the envelope of the VeLLOs is slightly larger than that of the starless cores but smaller than that of the protostellar cores. 
Here the infall speeds for protostellar cores by \citet{2001ApJ...562..770D} have been derived from the spectra obtained from high spatial ($\sim 2\arcsec$ resolution) observations that may give higher infall speeds compared to those from the spectra from single-dish observations. However, the infall speeds derived from \citet{2013A&A...558A.126M} using radiative transfer models for seven protostellar cores are found to be mostly between $V_{infall} \sim 0.3 - 0.7~\kms$ when the infalling area is probed by the spatial scale equivalent to that of the KVN. Thus it seems evident that the infall speed of the envelope of the VeLLOs is smaller than that of the protostellar cores.

It is worth noting that the infall speeds in the envelopes of the VeLLOs are mostly smaller than their sound speeds, which are primarily between $\sim 0.20 - 0.24~\kms$ depending on their temperatures, so called subsonic, while infall motions in four sources are equivalent to their sound speeds, trans-sonic. This is in contrast with the starless cores for which their infall speeds are all subsonic and the protostellar cores for which their infall speeds are mostly supersonic (the infall speeds for all protostellar cores are larger than the sound speeds $\sim 0.24~\kms$ at a temperature of 20 K.)

\begin{deluxetable*}{lccccllcccl}
\tabletypesize{\footnotesize}
\addtolength{\tabcolsep}{1pt}
\tablecaption{VeLLO-related physical properties \label{tab:tbl3}}
\tablehead{\colhead{Name} & \colhead{$\rm V_{infall}$} & \colhead{$\rm V_{ff}$} & \colhead{$\rm M_{env}$} & \colhead{$\rm M_{env}^{16\arcsec}$} & \colhead{$\dot{M}_{infall}$} & \colhead{$\dot{M}_{acc}^{CO}$} & \colhead{$T_d$} & \colhead{collapsing age} & \colhead{$M_{acc} ^{star}$} & \colhead{$M_{final} ^{star}$} \\
           \colhead{} & \colhead{$(\kms)$} & \colhead{$(\kms)$} & \colhead{$(\Msun)$} & \colhead{$(\Msun)$} & \colhead{$(\rm \Msun~yr^{-1})$} & \colhead{$(\rm \Msun~yr^{-1})$} & \colhead{(K)} & \colhead{(Myr)} & \colhead{$(\Msun)$} & \colhead{$(\Msun)$}}
\colnumbers
\startdata
        J032832  &  0.17$\pm$0.002  &  0.21$\pm$0.022 &  0.25$\pm$0.02 & 0.28$\pm$0.06  &  6.5$\pm$1.4 $\times 10^{-6}$    &  1.5$\pm$5.0 $\times 10^{-6*}$  & 12.5 & 0.11$\pm$0.01  & 0.17$\pm$0.06  & 0.22$\pm0.06$ \\  
        J033032  &  0.06$\pm$0.004  &  0.24$\pm$0.022 &  0.65$\pm$0.05 & 0.38$\pm$0.06  &  3.1$\pm$5.9 $\times 10^{-6}$    &  2.0$\pm$1.0 $\times 10^{-6}$   & 11.5 & 0.12$\pm$0.01  & 0.24$\pm$0.12  & 0.37$\pm0.13$ \\  
       J0418402  &  0.03$\pm$0.020  &  0.11$\pm$0.023 &  0.05$\pm$0.01 & 0.04$\pm$0.02  &  3.5$\pm$2.5 $\times 10^{-7}$    &  1.4$\pm$3.0 $\times 10^{-7*}$  & 12.5 & 0.06$\pm$0.01  & 0.008$\pm$0.002  & 0.02$\pm0.01$ \\  
     L1521F-IRS  &  0.03$\pm$0.002  &  0.24$\pm$0.019 &  3.06$\pm$0.15 & 0.19$\pm$0.03  &  1.7$\pm$2.9 $\times 10^{-6}$    &  1.5$\pm$1.0 $\times 10^{-8}$   & 11.1 & 0.06$\pm$0.01  & 0.0009$\pm$0.0006  & 0.61$\pm0.15$ \\  
       J0430149  &  0.12$\pm$0.002  &  0.10$\pm$0.012 &  0.03$\pm$0.01 & 0.03$\pm$0.01  &  1.1$\pm$2.5 $\times 10^{-6}$    &  5.1$\pm$1.8 $\times 10^{-7*}$  & 13.1 & 0.05$\pm$0.01  & 0.03$\pm$0.01  & 0.03$\pm0.01$ \\  
       L328-IRS  &  0.04$\pm$0.005  &  0.15$\pm$0.028 &  0.07$\pm$0.02 & 0.10$\pm$0.04  &  7.2$\pm$2.9 $\times 10^{-7}$    &  8.0$\pm$1.0 $\times 10^{-7}$   & 13.1 & 0.08$\pm$0.01  & 0.07$\pm$0.01  & 0.08$\pm0.02$ \\  
       J1830144  &  0.17$\pm$0.026  &  0.23$\pm$0.035 &  1.69$\pm$0.34 & 0.58$\pm$0.13  &  8.1$\pm$2.7 $\times 10^{-6}$    &  1.4$\pm$5.0 $\times 10^{-6*}$  & 12.6 & 0.19$\pm$0.04  & 0.27$\pm$0.11  & 0.61$\pm0.35$ \\  
       J1830156  &  0.27$\pm$0.026  &  0.48$\pm$0.099 &  0.75$\pm$0.16 & 2.47$\pm$0.90  &  5.5$\pm$2.3 $\times 10^{-5}$    &  9.1$\pm$3.2 $\times 10^{-7*}$  & 12.9 & 0.19$\pm$0.04  & 0.17$\pm$0.07  & 0.66$\pm0.90^{**}$ \\
       J1830162  &  0.22$\pm$0.033  &  0.32$\pm$0.053 &  0.43$\pm$0.09 & 1.08$\pm$0.29  &  2.0$\pm$7.2 $\times 10^{-5}$    &  1.2$\pm$6.0 $\times 10^{-6*}$  & 12.5 & 0.19$\pm$0.04  & 0.22$\pm$0.12  & 0.44$\pm0.32^{**}$ \\ 
       J1832424  &  0.10$\pm$0.025  &  0.22$\pm$0.033 &  0.51$\pm$0.10 & 0.52$\pm$0.12  &  4.3$\pm$1.7 $\times 10^{-6}$    &  3.7$\pm$1.3 $\times 10^{-7*}$  & 13.8 & 0.18$\pm$0.04  & 0.07$\pm$0.03  & 0.17$\pm0.11$ \\  
      L1148-IRS  &  0.11$\pm$0.038  &  0.15$\pm$0.021 &  0.32$\pm$0.02 & 0.19$\pm$0.05  &  2.3$\pm$1.0 $\times 10^{-6}$    &  8.0$\pm$1.0 $\times 10^{-7}$   & 12.3 & 0.14$\pm$0.01  & 0.11$\pm$0.02  & 0.18$\pm0.02$ \\  
       J2102212  &  0.22$\pm$0.023  &  0.29$\pm$0.021 &  0.17$\pm$0.02 & 0.67$\pm$0.09  &  1.7$\pm$2.9 $\times 10^{-5}$    &  1.2$\pm$1.0 $\times 10^{-5}$   & 12.0 & 0.14$\pm$0.01  & 1.64$\pm$0.16  & 1.78$\pm0.19^{**}$ \\ 
       J2102273  &  0.16$\pm$0.011  &  0.27$\pm$0.018 &  0.11$\pm$0.01 & 0.59$\pm$0.07  &  1.0$\pm$1.5 $\times 10^{-5}$    &  \nodata                        & 12.2 & 0.14$\pm$0.01  & \nodata        & \nodata \\  
       J2144570  &  0.16$\pm$0.027  &  0.30$\pm$0.036 &  1.96$\pm$0.11 & 1.53$\pm$0.37  &  1.2$\pm$3.7 $\times 10^{-5}$    &  \nodata                        & 12.5 & 0.31$\pm$0.01  & \nodata        & \nodata \\  
       J2229594  &  0.24$\pm$0.029  &  0.23$\pm$0.026 &  0.32$\pm$0.02 & 0.41$\pm$0.09  &  1.1$\pm$2.8 $\times 10^{-5}$    &  2.1$\pm$7.0 $\times 10^{-7*}$  & 12.1 & 0.14$\pm$0.01  & 0.03$\pm$0.01  & 0.11$\pm0.09^{**}$ \\
\enddata
\tablecomments{Col.(1) lists the source name. Col.(2) gives the infall speed derived from HILL5 model fit for HCN (1-0) line. Its error is given by 1 $\sigma$ uncertainty of the speed obtained from the HILL5 model fits for 1000 artificial spectra produced by the best fit parameters and observational rms. Col.(3) indicates free-fall speeds for the envelope by the area encompassed by the radius ($\sim 16\arcsec$) of the KVN telescope. Col.(4) gives the envelope mass which is based on the photometry with CSAR \citep{2016ApJS..225...26K}. Col.(5) lists the envelope mass based on the aperture photometry of 250 $\micron$ {\it Herschel} flux for the area encompassed by the radius ($\sim 16\arcsec$) of the KVN telescope. The uncertainty was calculated based on the photometry error. Col.(6) and Col.(7) provide the mass infall rate derived from the infall speed and the mass accretion rate inferred from CO outflow \citep{2019ApJS..240...18K}, respectively. The mass accretion rates marked with $*$ were derived from this study. Col.(8) lists the dust temperature and its $\sim 5\%$ uncertainty \citep{2014AA...562A.138R}. Col.(9) indicates the period for the expansion sound wave in an isothermal sphere to travel to its radius which is half of our observing beam size in here\citep{1977ApJ...214..488S}. Col.(10) lists the possible central mass of the VeLLOs to acquire with the mass accretion rates during the collapsing age given in Col.(9). Col.(11) indicates a possible final mass of the VeLLO to achieve, a sum of the possible central mass (in Col.(9)) plus the portion ($\sim0.2$) of the envelope mass (in Col.(4)). The stellar masses ($M_{star}$) with $**$ were calculated using the envelope mass given in Col.(5). The ellipsis symbols ({$\cdots$}) are to indicate ``No outflow observation'' and thus no further estimation of its corresponding physical quantities.}
\end{deluxetable*}

What would be the physical origin for the infall motions derived from the line asymmetry observed toward the envelopes where the VeLLOs are forming? These motions could be gravitational or non-gravitational such as converging flows, core accretion, or oscillatory motions.
For this discussion, we simply derive a mean gravitational free-fall speed in a pressure-free dense core, which is given by $V_{ff} = (2/\pi)(2GM_{env}/R)^{1/2}$ by dividing $R$ with its free-fall time scale of the core $\tau_{ff} = (3\pi/32G \rho_0)^{1/2}$ where $M_{env}$ is the mass of the core and $\rho_0$ is its mean density. 
Here we derived $M_{env}$ by using 250 $\micron$ continuum emission obtained from the {\it Herschel} SPIRE for the area encompassed by a radius of 16$\arcsec$ which corresponds to the radius of the beam size of the KVN telescope.
Considering an optically thin approximation, the mass can be obtained by $M_{env}=d^2F_\nu/(\kappa_\nu B_\nu(T))$. Here, $d$ is the distance to the source, $F_{250}$ is the flux at 250 $\micron$ from the {\it Herschel} SPIRE, $\kappa_\nu$ is the opacity, and $B_\nu(T)$ is the Planck function \citep{2016ApJS..225...26K} at the temperature T. 
The T here is the dust temperature adopted from the {\it Herschel} Gould Belt survey archive \citep{2010A&A...518L.102A} and found to be between 11.1--13.8 K for infalling cores with the VeLLOs. For the opacity, the value of 0.19 $\rm cm^2/g$ was adopted from \citet{1994A&A...291..943O} at the wavelength of 250 $\micron$. At this opacity value, the optical depth at 250 $\micron$ is found to be far less than 1, and thus our mass estimation using the optically thin approximation is thought to be reasonable.

\begin{figure}[htb!]
  \centering
  \includegraphics[scale=0.55]{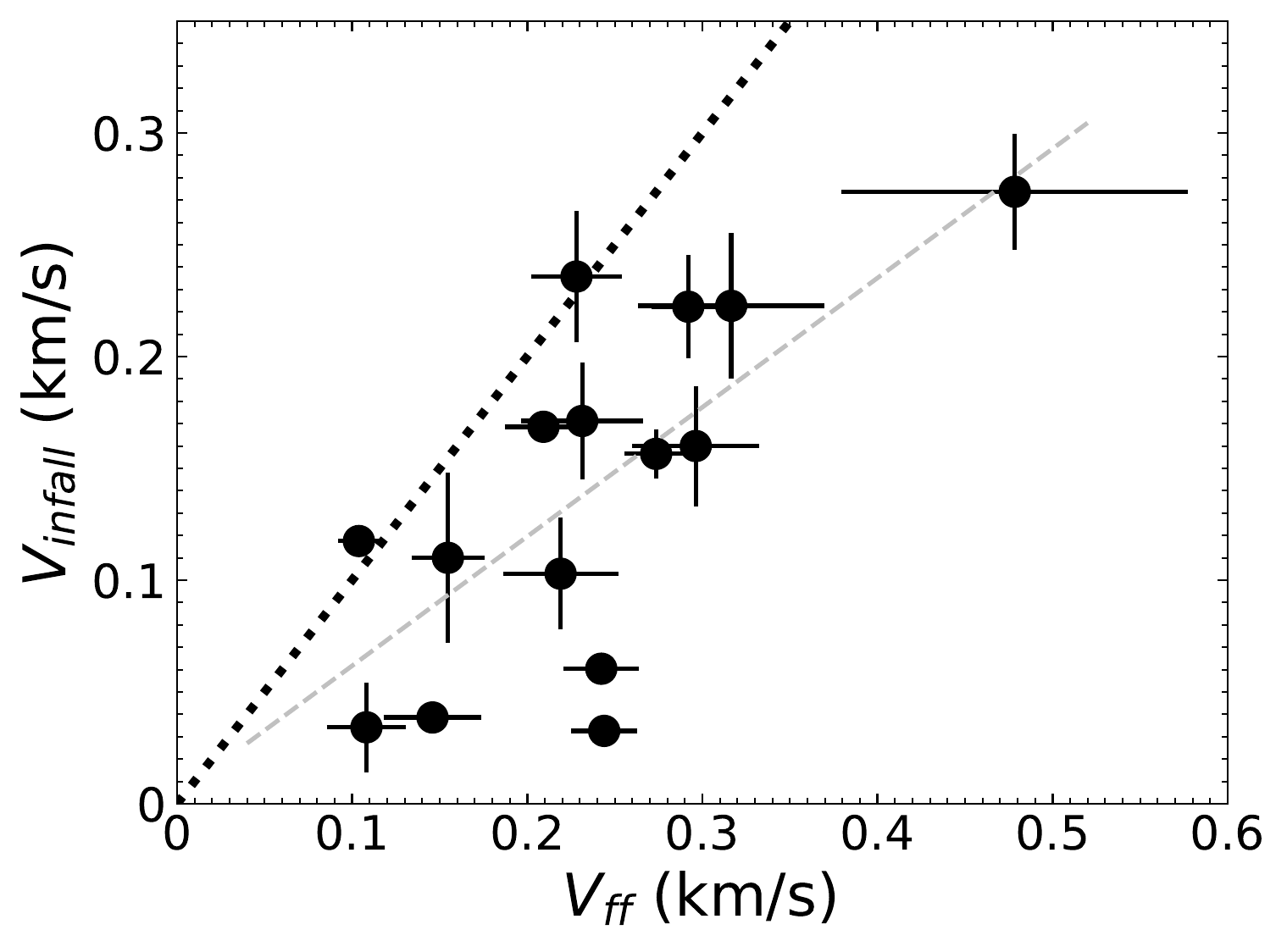}
  \caption{Comparison of the infall velocities for 15 cores with VeLLOs with their free-fall velocities. The black dotted line indicates the positions where $V_{infall}$ is equal to $V_{ff}$. The grey dashed line is a line for the linear least-squares fit to the data. Its slope and Pearson correlation coefficient are 0.60$\pm$0.17 and 0.69, respectively. \label{fig:vff_vin}}
\end{figure}

Figure \ref{fig:vff_vin} compares the infall speeds of our targets with the mean free-fall speeds of the pressure-free theoretical cores whose masses are the same as those of our observing targets, indicating that the motions derived from the line profiles for most of our targets seem to be well correlated with a simple model of gravitational contraction with the correlation coefficient of about 0.7.
In fact, it is noted that the infall speeds do not exactly correspond to the same values of the free-fall speeds, but to about half of those values. Considering the large uncertainties in both the data and the applicability of the simple model we have used, however, $V_{infall}$ and $V_{ff}$ are thought to be well correlated.

For example, for a line profile having a well-resolved dip or shoulder, a peak brightness temperature greater than 1 K and a signal-to-noise ratio greater than 30, the infall speed estimated using the HILL5 model is uncertain by 0.02 $\kms$ \citep{2005ApJ...620..800D}. This becomes an additional source of uncertainty on top of the imprecision, or the variation ($\rm \sim 0.05~\kms$) due to the noise quoted for the lower SNR sources in this paper. 
We also note that the free-fall time is a time scale which is only useful as a guide and hence cannot be used for an exact prediction, since the initial conditions such as a uniform density, pressure-free sphere collapsing from a rest condition with no rotation and no magnetic field are not realistic.
Given the large uncertainties in the measurements of the infall speeds and and the unrealistic simplicity of the collapse model in calculating the free-fall speeds, it seems reasonable to say that as a group they show significant consistency between the infall speeds and about half the free-fall speeds.
 
Therefore, our results indicate that the infall motions inferred by the spectral infall asymmetry in the envelopes with VeLLOs are more likely gravitational, favoring models of gravitational infall motion over models of non-gravitational motion such as converging flows (e.g., \citealt{2009ApJ...699..230G}), or oscillation motions without infall (e.g., \citealt{2010ApJ...721..493B}) as it is unlikely that the models other than gravitational infall models can produce this relation (e.g., \citealt{2019ApJ...871..134S}).

Alternatively, it may be possible that the collapsing radius is far smaller than the telescope beam size for our observations ($\sim 12,000$ au for the most distant target), making the infall speed to be lower than the speed estimated by the gravitational collapse of the beam area. 
It is noted that there are two sources (J0430149 and J2229594) for which $V_{ff}$s are about the same as $V_{infall}$s. 
For those two sources, the infall motions inferred by the spectral infall asymmetry in the envelopes may be more likely fully gravitational. If then, it may be possible that their collapsing radius may be close to 2300 au or 5600 au, which is the linear radius of the HPBW of the KVN telescope at the distances of two VeLLOs. On the other hand, there are other targets whose infall velocities are smaller than the free fall velocity at the distances comparable to those of these two sources. This indicates that the collapsing radius of the core may not be uniquely determined but variable depending on the physical status of the core.
Any further quantitative discussion in comparison of our observations with results predicted by the various models dealing with gas infall motions is beyond the scope of this paper and we leave this issue for future study.

\subsection{Mass infall rates and mass accretion rates toward the VeLLOs, and their implication}
In this section, we consider the mass infall rates from the envelopes of the VeLLOs and the mass accretion rates toward the VeLLOs as useful physical quantities to help to understand the physical processes involved in the formation of the VeLLOs from their parent envelopes.
The mass infall rates were derived using the infall speeds obtained from our model fitting, with $\dot{M}_{infall} = 4 \pi R^2_{in} \rho_0 V_{infall}$, where $\rho_0$ is the mean density of the envelope given by $M_{env}/(4/3 \pi R^3_{in})$ and $R_{in}$ is the half of FWHM (16$\arcsec$) of our observing KVN telescope. It is assumed that the envelope has a uniform density $\rho$ and the infall is spherically symmetric with a speed of $V_{infall}$. Estimated mass infall rates range from $10^{-7}$ to $10^{-5} \Msun~yr^{-1}$, with the median value of $\rm 3.4 \pm 1.5 \times 10^{-6}\Msun~yr^{-1}$ (Table \ref{tab:tbl3}).
The values of $\kappa$ can be a significant contributor to the uncertainties in the mass and then the mass infall rate, which may vary within a factor of 2 due to the uncertainty in the value of $\kappa$ \citep{1994A&A...291..943O}.
We note that line profiles showing an infall asymmetry can be affected by an outflow activity in their shapes if it is strong enough, and thus the infall speed and the mass infall rate that we derived can be affected. We found that 13 out of 15 sources for which infall speeds are derived from their infall asymmetric profiles have possible outflow wings from the CO observations by \citet{2019ApJS..240...18K}, but they are found to be all weak enough not to significantly affect the derived values of infall speed and the mass infall rate. 

The mass accretion rate can be obtained by using CO observations around the outflow regions. \citet{2019ApJS..240...18K} made a single pointing and/or a mapping survey of 68 VeLLOs in CO lines, finding the evidence for outflows over 16 dense cores having the VeLLOs.

We found that there are 5 outflow sources identified from the CO mapping survey for which the mass infall rates were estimated in this study. For these sources, we used the mass accretion rates given by \citet{2019ApJS..240...18K}. We also found that there are 8 more outflow sources identified from the single pointing CO observations for which the mass infall rates were estimated in this study. for these sources we calculated the mass accretion rates using the equation given by \citet{2019ApJS..240...18K} with the same parameters assumed in their study; $\dot{M}_{\rm acc} = \frac{1}{f_{\rm ent}} \frac{\dot{M}_{\rm acc}}{\dot{M}_{\rm w}}\frac{1}{v_{\rm w}}F {{\rm sin ~\it i}\over {\rm cos^2 ~\it i}}$, where $f_{\rm ent}$ is the entrainment efficiency (0.25), $\frac{\dot{M}_{\rm acc}}{\dot{M}_{\rm w}}$ the ratio of the mass accretion rate to the mass loss rate in the jet/wind (10), $v_{\rm w}$ is the velocity of the jet/wind (150 km s$^{-1}$), $F$ is the outflow force, and $i$ is the inclination angle of the outflow (57.$^{\circ}$3 as an average value among its random outflow orientations). Thus, we were able to have both the mass accretion rates and mass infall rates for 13 sources in total for our discussion.

\begin{figure}[htb!]
  \centering
  \includegraphics[scale=0.8]{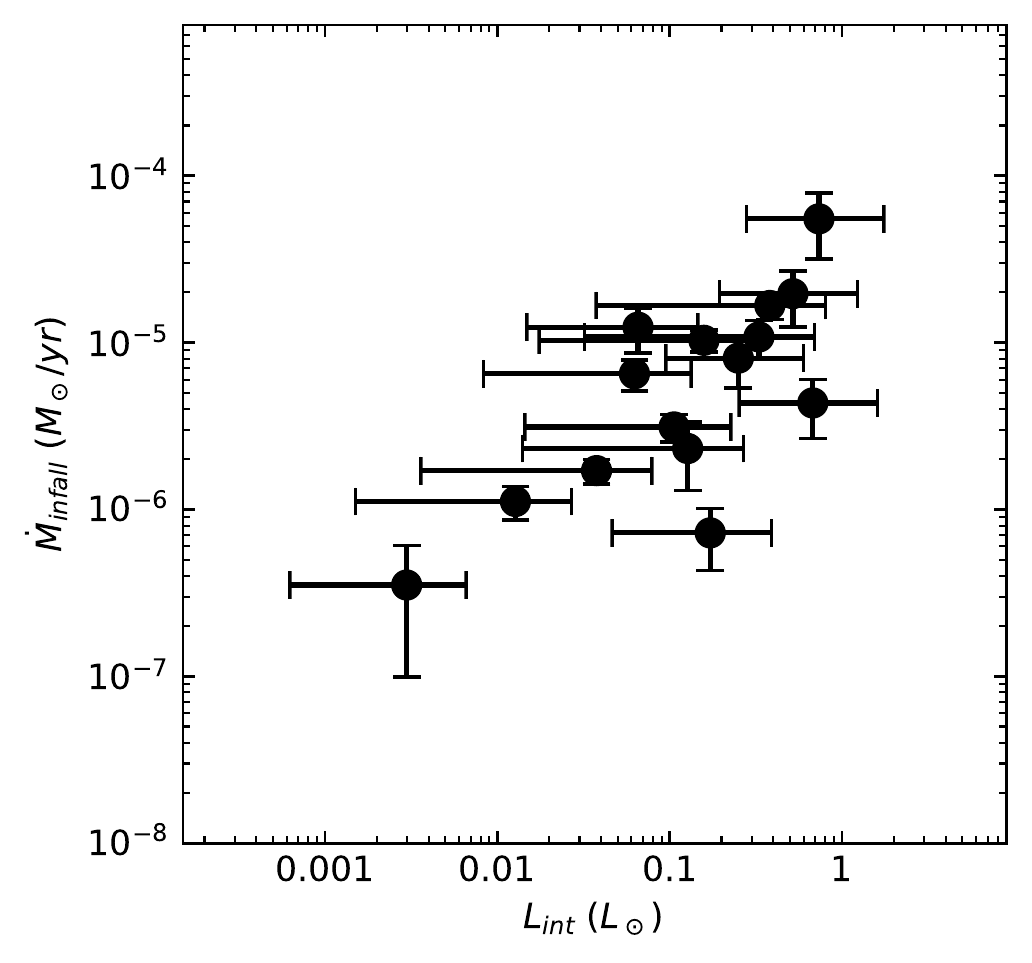}
  \caption{Mass infall rates toward the VeLLOs versus their internal luminosities. The Pearson correlation coefficient of the linear least-squares fit to the data is 0.70. \label{fig:acc_infall_Lint}}
\end{figure}

We investigated how the mass infall rates are related to the internal luminosities of the VeLLOs to help to understand the nature of the infalling motions found toward the VeLLOs.
A significant part of the luminosity of a protostar is believed to be due to the gravitational accretion of material onto the protostar, which may be related to the infalling motions in its envelope (e.g., \citealt{1987ARA&A..25...23S}).
Therefore, the mass infall rates of the VeLLOs are expected to be somehow correlated to their internal luminosities. Figure \ref{fig:acc_infall_Lint} indicates that these physical quantities seem to have a fairly good relationship with the internal luminosities. 

The mass infall rates toward the VeLLOs that we derived can be compared with what the collapse models predict. The mass infall rate based on the inside-out collapse model \citep{1977ApJ...214..488S}, $\dot{M} = 0.975 a^3/G$, where $a$ is the sound speed, is estimated to be $1.4-1.9 \times 10^{-6}~\Msun~yr^{-1}$ for isothermal cores whose temperatures are given as dust temperatures listed in Table \ref{tab:tbl3}, being consistent with the mass infall rates that we obtained for our targets within an order of magnitude range.
 
The good correlation found between the mass infall rates and the internal luminosities, and the consistency of the mass infall rates with the values predicted with the inside-out collapse model again indicate that the infall asymmetry observed in the envelope of 0.04 pc (32$\arcsec$) extent may have resulted in due to the gravitational infall motion.

\begin{figure}[htb!]
  \centering
  \includegraphics[scale=0.8]{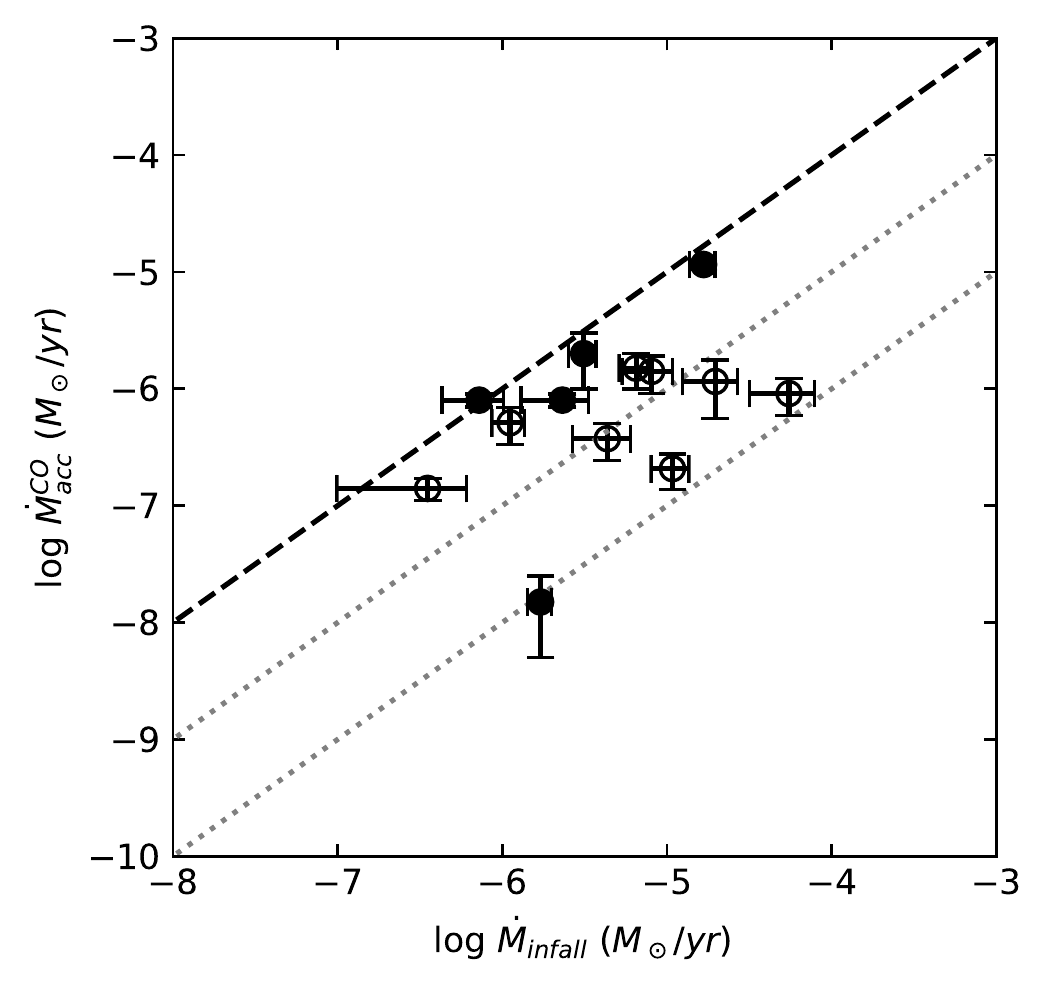}
  \caption{Mass infall rates versus the mass accretion rates toward the VeLLOs. The filled and open circles indicate the mass accretion rates obtained from CO observations in a mapping mode and a single pointing mode in CO line \citep{2019ApJS..240...18K}, respectively. The dashed and dotted lines are the positions where the ratios of $\dot{M}_{acc}^{CO}$ / $\dot{M}_{infall}$ are 1.0, 0.1 and 0.01, respectively. \label{fig:acc_infall}}
\end{figure}
Next we compare the mass infall rates with the mass accretion rates for our samples in Figure \ref{fig:acc_infall} to see how they are related. In this figure, the mass infall rates of almost half samples are similar the values of the mass accretion rates, while the mass infall rates for the rest of samples are larger than the mass accretion rates.
We speculate that for the case of the sources for which the mass accretion rates are comparable with the mass infall rates, the infalling material is expected to reach the central source through the normal accretion processes. However, for most sources for which the mass accretion rates are one order of magnitude lower than the mass infall rates, it may be possible that they are currently accreting only a small portion of the infalling material in a quiescent manner while most of the material is getting accumulated for future episodic accretion events. It may also be possible that the latter group's sources may be accreting only a fraction of the infalling material from the envelopes associated with them, while most of the infalling material may get ejected in the form of outflow winds before it reaches the VeLLOs.

\subsection{Implication on the identity of the VeLLOs}
Study of the VeLLOs are important to understand how faint protostars collect material from their envelopes and yet remain faint. The study is also useful to identify the precursors of brown dwarfs which in turn would help us to understand the processes involved in their formation and evolution.
In fact, by inferring the final masses that our target sources can possibly achieve, we can identify some of the potential precursors of brown dwarfs among the sources studied here.
For that, first, we calculated the possible central mass of the VeLLOs by assuming that their central sources are acquiring the mass with the mass accretion rates during the duration of isothermal inside-out collapse by \citet{1977ApJ...214..488S}, which can be derived as the period for the expansion sound wave to travel to the radius of our observing beam size.
Then assuming that the central source will collect the additional mass from its envelope with a core-to-star formation efficiency of $\sim 0.2$ (e.g., \citealt{2009ApJS..181..321E}), the final masses of the VeLLOs that could be collected during the whole duration of the star/brown dwarf formation can be inferred as a sum of the possible central mass plus the portion ($\sim 0.2$) of the envelope mass. Those inferred values are listed in the last column of Table \ref{tab:tbl3}. We find that most of our targets would have a stellar mass while at least two (J0418402 and J0430149) of our targets would have a brown dwarf mass within the uncertainty. It is interesting to note that the final masses of these two sources will be in the brown-dwarf mass regime, even though we assume that the entire infalling material finally arrive at the central sources. Further studies of these two sources in future may help us to better test the proto-brown dwarf nature of these sources.
In conclusion, it is inferred that the VeLLOs studied here are mostly faint protostars and proto-brown dwarfs (at least in two cases) and thus may be extremely important objects to study the early formation processes of low-mass protostars or proto-brown dwarfs.

\section{Summary}
We conducted single pointing observations toward the dense cores harboring VeLLOs in optically thick (HCN 1-0) and thin ($\NH$ 1-0) molecular lines using KVN and Mopra telescopes to study their inward motions and and to characterize their properties. Our results are summarized as follows;

(1) Out of 95 sources observed, we detected 41 sources in both the lines and analyzed the profile shapes in a quantitative manner using the normalized velocity difference ($\delta V$) between two tracers, finding that the majority (21) of our sources show the typical ``blue'' asymmetry in the HCN 1-0 profile, six sources show the ``red'' asymmetric features, and 6 sources show a ``mixture of both blue and red'' asymmetry.
We found that the $\delta V$ distribution of the dense cores with the VeLLOs is highly skewed to the blue, indicating the infalling motions in their envelopes are dominant. This is found to be more similar to that of the starless cores than that of the normal protostars, implying that the influence of the VeLLO formation activity on its envelope is minimal on a scale traced by the single dish telescopes.

(2) The distribution of line widths in the dense cores with the VeLLOs indicates that they are mostly in the regime where thermal motions are dominant, while a significant number (about 20 \%) of the sources are located in a dominant turbulence regime. This is in good agreement with the kinematic environment of the starless cores that are in a state of thermal infall unlike the protostellar cores dominated by the turbulent motion.

(3) We derived the infall speeds of the infall motions found in the envelopes of the VeLLOs and the related physical quantities by applying the HILL5 radiative transfer model to the HCN spectra of the sources showing infall signature. The resulting infall speeds for the sources are in the range $\rm 0.03 < V_{infall} < 0.3~\kms$ with the median value of $\rm 0.16~\kms$.

These values are found to be well correlated to about half of the gravitational free-fall speeds from pressure-free envelopes, indicating that the infall motions inferred from the spectral infall asymmetry in the envelopes with the VeLLOs are more likely gravitational considering the large uncertainties in both the data and the applicability of the simple collapse model.

Alternatively, these smaller values of the infall speeds may be due to the collapsing radii of the envelopes being smaller than our telescope beam size. Nonetheless, we found two sources (J0430149 and J2229594) for which $V_{ff}$s are about the same as $V_{infall}$s. If this indicates that the infall motions in these sources are from a fully gravitational origin, their collapsing radii are expected to be 2300 au or 5600 au.

(4) The mass infall rates $\rm \dot{M}_{infall}$ for our targets were estimated to be mostly of the order of $\rm \sim 10^{-6} \Msun~yr^{-1}$ with the median value of $\rm 3.4 \times 10^{-6} \Msun~yr^{-1}$ and compared among other physical properties such as their internal luminosities, and the mass infall rates ($1.4-1.9 \times 10^{-6}~\Msun~yr^{-1}$) by the inside-out collapse model. We found that there is a fairly good relation between the mass infall rates and the internal luminosities, and also the mass infall rates are consistent with the values predicted with the inside-out collapse model. This again indicates that the infall asymmetry observed toward the VeLLOs may have resulted in due to the gravitational infall motions in their envelopes.

(5) From the comparison of the mass infall rates for a sub-sample of targets with the corresponding mass accretion rates, it was found that in almost half of the sources the mass infall and mass accretion rates are comparable while in the rest the mass infall rates are an order of magnitude larger than their mass accretion rates. The sources in the former group are expected to accrete material continuously. On the other hand, those in the latter group are probably the sources where the envelope material is either at present quiescently accreting at a low rate expecting for their future episodic high accretion events or the sources where the envelope material may get ejected out in outflow activities and never reach the VeLLOs.

(6) Considering the mass infall rates and the mass accretion rates of our sources, the duration of isothermal inside-out collapse in the envelope of the telescope beam size, and the envelope masses, we estimated the final expected masses of the VeLLOs, finding that most of our targets would have a stellar mass while at least two (J0418402 and J0430149) of our targets would have a brown dwarf mass. This concludes that the VeLLOs can be either faint protostars or proto-brown dwarfs, confirming them to be extremely important objects to study the early formation processes of low-mass protostars or proto-brown dwarfs.

Future spectroscopic observations with facilities of high angular resolution and high sensitivity such as ALMA will be highly helpful to better test their true nature.

We thank an anonymous referee for valuable comments, which greatly helped to improve the paper.
This work was supported by Basic Science Research Program though the National Research Foundation of Korea (NRF) funded by the Ministry of Education, Science, and Technology (NRF-2017R1A6A3A01075724 and NRF-2019R1A2C1010851).
This research has made use of data from the Herschel Gould Belt survey (HGBS) project (http://gouldbelt-herschel.cea.fr). The HGBS is a Herschel Key Programme jointly carried out by SPIRE Specialist Astronomy Group 3 (SAG 3), scientists of several institutes in the PACS Consortium (CEA Saclay, INAF-IFSI Rome and INAF-Arcetri, KU Leuven, MPIA Heidelberg), and scientists of the Herschel Science Center (HSC).


\end{document}